\documentclass[11pt,a4paper]{article}

\usepackage{fullpage}

\usepackage[centertags]{amsmath}
\allowdisplaybreaks[1]
\usepackage{amsbsy}
\usepackage{amsfonts}
\usepackage{amssymb}
\usepackage{revsymb}

\usepackage{cite}

\usepackage{graphicx}% Include figure files
\usepackage{dcolumn}% Align table columns on decimal point

\usepackage{bm}% bold math
\usepackage{mathrsfs}

\usepackage{subfigure}

\renewcommand{\today}{\number\day\space\ifcase\month\or
January\or February\or March\or April\or May\or June\or
July\or August\or September\or October\or November\or December\fi
\space\number\year}

\newcommand{\sla}[1]{ \ensuremath{ \rlap{\hskip0.5pt/} {#1} } }
\newcommand{\vet}[1]{\ensuremath{\hskip-1pt\vec{\hskip1pt#1}}}

\newcommand{\chg}{\ensuremath{q}}
\newcommand{\mgm}{\ensuremath{\mu}}
\newcommand{\elm}{\ensuremath{\epsilon}}
\newcommand{\anm}{\ensuremath{a}}

\begin{document}

\begin{center}
\Large\bfseries
Electromagnetic Properties of Neutrinos
\\[0.5cm]
\large\normalfont
C. Broggini\ensuremath{^{(a)}},
C. Giunti\ensuremath{^{(b)}},
A. Studenikin\ensuremath{^{(c)}}
\\[0.5cm]
\normalsize\itshape
\setlength{\tabcolsep}{1pt}
\begin{tabular}{cl}
\ensuremath{(a)}
&
INFN, Sezione di Padova,
Via F. Marzolo 8, I--35131 Padova, Italy
\\[0.3cm]
\ensuremath{(b)}
&
INFN, Sezione di Torino,
Via P. Giuria 1, I--10125 Torino, Italy
\\[0.3cm]
\ensuremath{(c)}
&
Department of Theoretical Physics, Moscow State University, 119991 Moscow, Russia;
\\
&
Joint Institute for Nuclear Research, Dubna 141980, Moscow Region, Russia
\end{tabular}
\end{center}
\begin{abstract}
In this review\footnote{Invited review for the special issue of Advances in High Energy Physics on Neutrino Physics}
we discuss the main theoretical aspects and experimental effects of
neutrino electromagnetic properties.
We start with a general description of the electromagnetic form factors
of Dirac and Majorana neutrinos.
Then, we discuss the theory and phenomenology of the magnetic and electric dipole moments,
summarizing the experimental results and the theoretical predictions.
We discuss also the phenomenology of a neutrino charge radius and radiative decay.
Finally,
we describe the theory of neutrino spin and spin-flavor precession in a transverse magnetic field
and we summarize its phenomenological applications.
\begin{center}
%\texttt{\dayofweekname{\day}{\month}{\year} \ddmmyydate\today, \currenttime}
%\texttt{(\today)}
\end{center}
\end{abstract}

\tableofcontents

\section{Introduction}
\label{sec1}

The investigation of neutrino properties is one of the most active
fields of research in current high-energy physics.
Neutrinos are special particles, because they interact very weakly
and their masses are much smaller than those of the other fundamental fermions
(charged leptons and quarks).
In the Standard Model
neutrinos are massless and have only weak interactions.
However, the observation of neutrino oscillations by many experiments
(see
\cite{Giunti-Kim-2007,GonzalezGarcia:2007ib,Bilenky:2010zza,Xing:2011zza})
imply that neutrinos are massive and mixed.
Therefore,
the Standard Model must be extended to account for neutrino masses.
In many extensions of the Standard Model neutrinos acquire also electromagnetic properties through
quantum loops effects.
Hence,
the theoretical and experimental study of neutrino electromagnetic interactions
is a powerful tool in the search for the fundamental theory beyond the Standard Model.
Moreover,
the electromagnetic interactions of neutrinos can generate important effects,
especially in astrophysical environments,
where neutrinos
propagate for long distances in magnetic fields
both in vacuum and in matter.

In this paper we review the theory and phenomenology of neutrino
electromagnetic interactions.
After a derivation of all the possible types of electromagnetic interactions
of Dirac and Majorana neutrinos
we discuss their effects in terrestrial and astrophysical environments
and the corresponding experimental results.
In spite of many efforts in the search of neutrino
electromagnetic interactions,
up to now there is no positive experimental indication in favor of their existence.
However,
the existence of neutrino masses and mixing imply that
non-trivial neutrino electromagnetic properties
are plausible
and
experimentalists and theorists are eagerly looking for them.

In this review we use the notation and conventions in \cite{Giunti-Kim-2007}.
When we consider neutrino mixing,
we have the relation
\begin{equation}
\nu_{{\alpha}L} = \sum_{k=1}^{3} U_{\alpha k} \, \nu_{kL}
\qquad
(\alpha=e,\mu,\tau)
\label{mixing}
\end{equation}
between the left-handed components
of the three flavor neutrino fields
$\nu_{e}$,
$\nu_{\mu}$,
$\nu_{\tau}$
and the left-handed components of three
massive neutrino fields
$\nu_{k}$
with masses $m_{k}$
($k=1,2,3$).
The $3\times3$ mixing matrix $U$ is
unitary ($U^{\dagger}=U^{-1}$).

Neutrino electromagnetic properties are discussed in the books in
\cite{CWKim-book,Fukugita:2003en,Mohapatra:2004,Xing:2011zza},
and in the previous reviews in
\cite{Raffelt:1990yz,Raffelt:2000kp,Pulido:1992fb,Nowakowski:2004cv,Wong:2005pa,Giunti:2008ve,Studenikin:2008bd}.

The structure of this paper is as follows.
In Section~\ref{sec3}
we discuss the general form of the electromagnetic interaction of Dirac and Majorana neutrinos,
which is expressed in terms of form factors,
and the derivation of the form factors in gauge models.
In Section~\ref{sec4}
we discuss the phenomenology of the neutrino magnetic and electric dipole moments in laboratory experiments.
These are the most studied electromagnetic properties of neutrinos,
both experimentally and theoretically.
In Section~\ref{sec5}
we review the theory and experimental constraints on the neutrino charge radius.
In Section~\ref{sec6}
we discuss neutrino radiative decay and the astrophysical bounds on a neutrino magnetic moment
obtained from the study of plasmon decay in stars.
In Section~\ref{sec7}
we discuss neutrino spin and spin-flavor precession.
In conclusion,
in Section~\ref{sec8}
we summarize the status of our knowledge of
neutrino electromagnetic properties and we discuss the prospects for future research.

\section{Electromagnetic form factors}
\label{sec3}

The importance of neutrino electromagnetic properties
was first mentioned by Pauli in 1930, when he postulated the
existence of this particle and discussed the possibility that the
neutrino might have a magnetic moment.
Systematic theoretical
studies of neutrino electromagnetic properties started after
it was shown that in the extended Standard Model with
right-handed neutrinos the magnetic moment of a massive neutrino
is, in general, nonvanishing and that its value is determined by
the neutrino mass
\cite{Marciano:1977wx,Lee:1977tib,Fujikawa:1980yx,Petcov:1976ff,Pal:1981rm,Shrock:1982sc,Bilenky:1987ty}.

Neutrino electromagnetic properties are important
because they are directly connected to fundamentals of particle
physics.
For example, neutrino electromagnetic properties can be
used to distinguish Dirac and Majorana neutrinos (see
\cite{Schechter:1981hw,Shrock:1982sc,Pal:1981rm,Nieves:1981zt,Kayser:1982br,Kayser:1984ge})
and also as probes of new
physics that might exist beyond the Standard Model (see
\cite{Bell:2005kz,Bell:2006wi,NovalesSanchez:2008tn}).

\subsection{Dirac neutrinos}
\label{ff:Dirac}

\begin{figure}
\begin{center}
\begin{tabular}{ccc}
\begin{minipage}{0.3\linewidth}
\begin{center}
\subfigure[]{\label{A003}
\includegraphics*[width=0.8\linewidth]{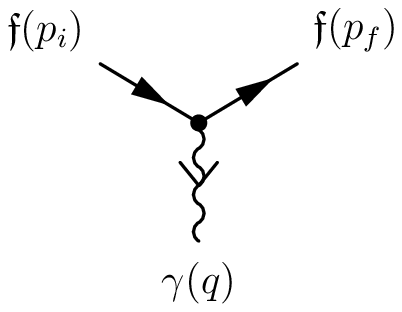}
}
\end{center}
\end{minipage}
&
\begin{minipage}{0.3\linewidth}
\begin{center}
\subfigure[]{\label{A004}
\includegraphics*[width=0.8\linewidth]{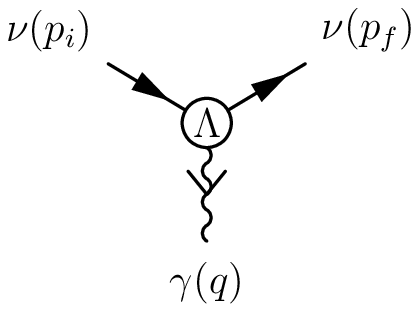}
}
\end{center}
\end{minipage}
&
\begin{minipage}{0.3\linewidth}
\begin{center}
\subfigure[]{\label{A005}
\includegraphics*[width=0.8\linewidth]{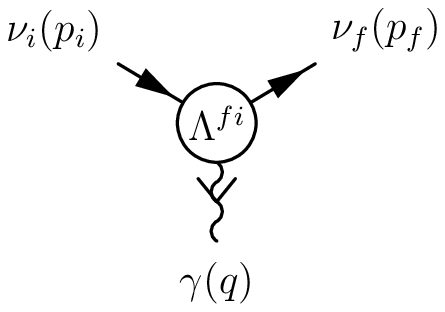}
}
\end{center}
\end{minipage}
\end{tabular}
\end{center}
\caption{ \label{A00345}
Tree-level coupling of a
charged fermion $\mathfrak{f}$ with a photon $\gamma$
\subref{A003},
effective coupling of a neutrino $\nu$ with a photon
\subref{A004}
and
effective coupling of neutrinos with a photon
taking into account possible transitions
between two different initial and final massive neutrinos $\nu_{i}$ and $\nu_{f}$
\subref{A005}.
}
\end{figure}

In the Standard Model,
the interaction of a fermionic field $\mathfrak{f}(x)$
with the electromagnetic field $A^{\mu}(x)$ is given by the interaction Hamiltonian
\begin{equation}
\mathcal{H}_{\text{em}}(x)
=
j_{\mu}(x) A^{\mu}(x)
=
\chg_{\mathfrak{f}} \overline{\mathfrak{f}}(x) \gamma_{\mu} \mathfrak{f}(x) A^{\mu}(x)
,
\label{A006}
\end{equation}
where
$\chg_{\mathfrak{f}}$ is the charge of the fermion $\mathfrak{f}$.
Figure~\ref{A003} shows the corresponding
tree-level Feynman diagram
(the photon $\gamma$ is the quantum of the electromagnetic field $A^{\mu}(x)$).

For neutrinos the electric charge is zero and there are no electromagnetic interactions at tree-level\footnote{
However, in some theories beyond the Standard Model
neutrinos can be millicharged particles
(see \cite{Giunti:2008ve}).
}.
However, such interactions can arise at the quantum level from loop diagrams at higher order of the perturbative expansion of the interaction.
In the one-photon approximation,
the electromagnetic interactions of a neutrino field $\nu(x)$
can be described
by the effective interaction Hamiltonian
\begin{equation}
\mathcal{H}_{\text{eff}}(x)
=
j_{\mu}^{\text{eff}}(x) A^{\mu}(x)
=
\overline{\nu}(x) \Lambda_{\mu} \nu(x) A^{\mu}(x)
,
\label{A007}
\end{equation}
where,
$j_{\mu}^{\text{eff}}(x)$
is the effective neutrino electromagnetic current four-vector
and
$\Lambda_{\mu}$
is a $4\times4$ matrix in spinor space which can contain space-time derivatives,
such that
$j_{\mu}^{\text{eff}}(x)$
transforms as a four-vector.
Since radiative corrections are generated by weak interactions which are not invariant under a parity transformation,
$j_{\mu}^{\text{eff}}(x)$ can be a sum of
polar and axial parts.
The corresponding
diagram for the interaction of a
neutrino with a photon is shown in Fig.~\ref{A004},
where the blob represents the quantum loop contributions.

We are interested in the neutrino part of the amplitude corresponding to the diagram in Fig.~\ref{A004},
which is given by the
matrix element
\begin{equation}
\langle \nu(p_{f},h_{f}) |
j_{\mu}^{\text{eff}}(x)
| \nu(p_{i},h_{i}) \rangle
,
\label{A015Z}
\end{equation}
where
$p_{i}$ ($p_{f}$)
and
$h_{i}$ ($h_{f}$)
are the four-momentum and helicity of the initial (final) neutrino.
Taking into account that
\begin{equation}
\partial^{\mu} j_{\mu}^{\text{eff}}(x)
=
i \left[ \mathcal{P}^{\mu}, j_{\mu}^{\text{eff}}(x) \right]
,
\label{A015}
\end{equation}
where
$\mathcal{P}^{\mu}$
is the four-momentum operator which generate translations,
the effective current can be written as
\begin{equation}
j_{\mu}^{\text{eff}}(x)
=
e^{i \mathcal{P} \cdot x}
j_{\mu}^{\text{eff}}(0)
e^{- i \mathcal{P} \cdot x}
.
\label{A015a}
\end{equation}
Since
$
\mathcal{P}^{\mu} | \nu(p) \rangle
=
p^{\mu} | \nu(p) \rangle
$,
we have
\begin{equation}
\langle \nu(p_{f}) |
j_{\mu}^{\text{eff}}(x)
| \nu(p_{i}) \rangle
=
e^{i (p_{f}-p_{i}) \cdot x}
\langle \nu(p_{f}) |
j_{\mu}^{\text{eff}}(0)
| \nu(p_{i}) \rangle
,
\label{A015b}
\end{equation}
where we suppressed for simplicity the helicity labels
which are not of immediate relevance.
Here we see that the unknown quantity which determines the neutrino-photon interaction is
$
\langle \nu(p_{f}) |
j_{\mu}^{\text{eff}}(0)
| \nu(p_{i}) \rangle
$.
Considering that the incoming and outgoing
neutrinos are free particles which are described by
free Dirac fields with the standard Fourier expansion in Eq.~(2.139) of \cite{Giunti-Kim-2007},
we have
\begin{equation}
\langle \nu(p_{f}) |
j_{\mu}^{\text{eff}}(0)
| \nu(p_{i}) \rangle
=
\overline{u}(p_{f})
\Lambda_{\mu}(p_{f},p_{i})
u(p_{i})
.
\label{matr_elem}
\end{equation}
The electromagnetic properties of neutrinos are embodied by
$\Lambda_{\mu}(p_{f},p_{i})$,
which is a matrix in spinor space and can be decomposed
in terms of linearly independent products
of Dirac $\gamma$ matrices and the available kinematical four-vectors
$p_{i}$
and
$p_{f}$.
The most general decomposition can be written as
(see \cite{Nowakowski:2004cv})
\begin{equation}
\Lambda_{\mu}(p_{f},p_{i})
=
f_{1}(q^{2}) q_\mu
+
f_{2}(q^{2}) q_\mu \gamma_{5}
+
f_{3}(q^{2}) \gamma_\mu
+
f_{4}(q^{2}) \gamma_\mu \gamma_{5}
+
f_{5}(q^{2}) \sigma_{\mu\nu} q^{\nu}
+
f_{6}(q^{2}) \epsilon_{\mu\nu\rho\gamma} q^{\nu} \sigma^{\rho\gamma}
,
\label{A011}
\end{equation}
where
$f_{k}(q^{2})$
are six Lorentz-invariant form factors
($k=1,\ldots,6$)
and
$q$ is the four-momentum of the photon,
which is given by
\begin{equation}
q = p_{i} - p_{f}
,
\label{A009}
\end{equation}
from energy-momentum conservation.
Notice that the form factors depend only on $q^{2}$,
which is the only available Lorentz-invariant
kinematical quantity,
since
$(p_{i}+p_{f})^{2} = 4 m^{2} - q^{2}$.
Therefore,
$\Lambda_{\mu}(p_{f},p_{i})$
depends only on
$q$
and from now on we will denote it as
$\Lambda_{\mu}(q)$.

Since the Hamiltonian and the electromagnetic field are Hermitian
($\mathcal{H}_{\text{eff}}^{\dagger}=\mathcal{H}_{\text{eff}}$
and
$A^{\mu\dagger}=A^{\mu}$),
the effective current must be Hermitian,
$j_{\mu}^{\text{eff}\dagger}=j_{\mu}^{\text{eff}}$.
Hence,
we have
\begin{equation}
\langle \nu(p_{f}) |
j_{\mu}^{\text{eff}}(0)
| \nu(p_{i}) \rangle
=
\langle \nu(p_{i}) |
j_{\mu}^{\text{eff}}(0)
| \nu(p_{f}) \rangle^{*}
,
\label{A012a}
\end{equation}
which leads to
\begin{equation}
\Lambda_{\mu}(q)
=
\gamma^{0} \Lambda_{\mu}^{\dagger}(-q) \gamma^{0}
.
\label{A012}
\end{equation}
This constraint implies that
\begin{equation}
f_{2}
,
\quad
f_{3}
,
\quad
f_{4}
\quad
\text{are real}
,
\label{A013}
\end{equation}
and
\begin{equation}
f_{1}
,
\quad
f_{5}
,
\quad
f_{6}
\quad
\text{are imaginary}
.
\label{A014}
\end{equation}

The number of independent form factors can be reduced by imposing current conservation,
$\partial^{\mu} j_{\mu}^{\text{eff}}(x) = 0$,
which is required by gauge invariance
(i.e.
invariance of $\mathcal{H}_{\text{eff}}(x)$ under the transformation
$A^{\mu}(x) \to A^{\mu}(x) + \partial^{\mu} \varphi(x)$
for any $\varphi(x)$,
which leaves invariant the electromagnetic tensor
$F^{\mu\nu} = \partial^{\mu} A^{\nu} - \partial^{\nu} A^{\mu}$).
Using Eq.~(\ref{A015}),
current conservation implies that
\begin{equation}
\langle \nu(p_{f}) |
\left[ \mathcal{P}^{\mu}, j_{\mu}^{\text{eff}}(0) \right]
| \nu(p_{i}) \rangle
=
0
.
\label{A016}
\end{equation}
Hence,
in momentum space we have the constraint
\begin{equation}
q^{\mu}
\,
\overline{u}(p_{f})
\Lambda_{\mu}(q)
u(p_{i})
=
0
,
\label{A017}
\end{equation}
which implies that
\begin{equation}
f_{1}(q^{2}) q^{2}
+
f_{2}(q^{2}) q^{2} \gamma_{5}
+
2 m f_{4}(q^{2}) \gamma_{5}
=
0
.
\label{A018}
\end{equation}
Since $\gamma_{5}$ and the unity matrix are linearly independent,
we obtain the constraints
\begin{equation}
f_{1}(q^{2}) = 0
,
\quad
f_{4}(q^{2})
=
- f_{2}(q^{2}) q^{2} / 2 m
.
\label{A019}
\end{equation}
Therefore, in the most general case consistent with Lorentz and
electromagnetic gauge invariance, the vertex function is defined in terms of
four form factors \cite{Nieves:1981zt,Kayser:1982br,Kayser:1984ge},
\begin{equation}
\Lambda_{\mu}(q)
=
f_{Q}(q^{2}) \gamma_{\mu}
-
f_{M}(q^{2}) i \sigma_{\mu\nu} q^{\nu}
+
f_{E}(q^{2}) \sigma_{\mu\nu} q^{\nu} \gamma_{5}
+
f_{A}(q^{2}) (q^{2} \gamma_{\mu} - q_{\mu} \sla{q}) \gamma_{5}
.
\label{vert_func}
\end{equation}
where
$f_{Q} = f_{3}$,
$f_{M} = i f_{5}$,
$f_{E} = - 2 i f_{6}$ and
$f_{A} = - f_{2} / 2m$
are the real
charge, dipole magnetic and electric, and anapole neutrino form factors.
For the coupling with a real photon ($q^{2}=0$)
\begin{equation}
f_{Q}(0) = \chg
,
\quad
f_{M}(0) = \mgm
,
\quad
f_{E}(0) = \elm
,
\quad
f_{A}(0) = \anm
,
\label{A020a}
\end{equation}
where
$\chg$,
$\mgm$,
$\elm$ and
$\anm$
are, respectively,
the neutrino charge, magnetic moment, electric moment and anapole moment.
Although above we stated that $\chg=0$,
here we did not enforce this equality
because
in some theories beyond the Standard Model
neutrinos can be millicharged particles
(see \cite{Giunti:2008ve}).

Now it is interesting to study the properties of $\mathcal{H}_{\text{eff}}(x)$ under a CP transformation,
in order to find which of the terms in Eq.~(\ref{vert_func}) violate CP.
Let us consider the active CP transformation
\begin{equation}
\mathsf{U}_{\text{CP}}
\nu(x)
\mathsf{U}_{\text{CP}}^{\dagger}
=
\xi^{\text{CP}} \gamma^{0} \mathcal{C} \overline{\nu}^{T}(x_{\text{P}})
,
\label{ACP}
\end{equation}
where
$\xi^{\text{CP}}$ is a phase,
$\mathcal{C}$ is the charge-conjugation matrix
(such that
$
\mathcal{C}
\gamma_{\mu}^{T}
\mathcal{C}^{-1}
=
- \gamma_{\mu}
$,
$
\mathcal{C}^{\dagger} = \mathcal{C}^{-1}
$
and
$
\mathcal{C}^{T} = - \mathcal{C}
$),
and
$x^{\mu}_{\text{P}}=x_{\mu}$.
For the Standard Model electric current $j_{\mu}(x)$ in Eq.~(\ref{A006}) we have
\begin{equation}
j_{\mu}(x)
\xrightarrow{\;\text{CP}\;}
\mathsf{U}_{\text{CP}}
j_{\mu}(x)
\mathsf{U}_{\text{CP}}^{\dagger}
=
-
j^{\mu}(x_{\text{P}})
.
\label{A021a}
\end{equation}
Hence,
the Standard Model electromagnetic interaction Hamiltonian $\mathcal{H}_{\text{em}}(x)$
is left invariant by\footnote{
The transformation
$x \to x_{\text{P}}$
is irrelevant since all amplitudes are obtained by integrating over
$d^4x$.
}
\begin{equation}
A_{\mu}(x)
\xrightarrow{\;\text{CP}\;}
- A^{\mu}(x_{\text{P}})
.
\label{A021}
\end{equation}
CP is conserved in neutrino electromagnetic interactions
(in the one-photon approximation)
if $j_{\mu}^{\text{eff}}(x)$
transforms as $j_{\mu}(x)$:
\begin{equation}
\text{CP}
\quad
\Longleftrightarrow
\quad
\mathsf{U}_{\text{CP}}
j_{\mu}^{\text{eff}}(x)
\mathsf{U}_{\text{CP}}^{\dagger}
=
-
j^{\mu}_{\text{eff}}(x_{\text{P}})
.
\label{A022a}
\end{equation}
For the matrix element (\ref{matr_elem})
we obtain
\begin{equation}
\text{CP}
\quad
\Longleftrightarrow
\quad
\Lambda_{\mu}(q)
\xrightarrow{\;\text{CP}\;}
- \Lambda^{\mu}(q)
.
\label{A022}
\end{equation}
One can find that under a CP transformation we have
\begin{equation}
\Lambda_{\mu}(q)
\xrightarrow{\;\text{CP}\;}
\gamma^{0}
\mathcal{C}
\Lambda_{\mu}^{T}(q_{\text{P}})
\mathcal{C}^{\dagger}
\gamma^{0}
,
\label{A023}
\end{equation}
with
$q^{\mu}_{\text{P}}=q_{\mu}$.
Using the form factor expansion in Eq.~(\ref{vert_func}),
we obtain
\begin{equation}
\Lambda_{\mu}(q)
\xrightarrow{\;\text{CP}\;}
-
\left[
f_{Q}(q^{2}) \gamma^{\mu}
-
f_{M}(q^{2}) i \sigma^{\mu\nu} q_{\nu}
-
f_{E}(q^{2}) \sigma^{\mu\nu} q_{\nu} \gamma_{5}
+
f_{A}(q^{2}) (q^{2} \gamma^{\mu} - q^{\mu} \sla{q}) \gamma_{5}
\right]
.
\label{A024}
\end{equation}
Therefore,
only the electric dipole form factor violates CP:
\begin{equation}
\text{CP}
\quad
\Longleftrightarrow
\quad
f_{E}(q^{2}) = 0
.
\label{A025}
\end{equation}

So far,
we have considered only one massive neutrino field $\nu(x)$,
but according to the mixing relation (\ref{mixing}),
the three flavor neutrino fields
$\nu_{e}$,
$\nu_{\mu}$,
$\nu_{\tau}$
are unitary linear combinations of three massive neutrinos
$\nu_{k}$ ($k=1,2,3$).
Therefore, we must generalize the discussion to the case of more than one massive neutrino field.
The effective electromagnetic interaction Hamiltonian in Eq.~(\ref{A007})
is generalized to
\begin{equation}
\mathcal{H}_{\text{eff}}(x)
=
j_{\mu}^{\text{eff}}(x) A^{\mu}(x)
=
\sum_{k,j=1}^{3}
\overline{\nu_{k}}(x) \Lambda^{kj}_{\mu} \nu_{j}(x) A^{\mu}(x)
,
\label{A026}
\end{equation}
where we take into account possible transitions between different massive neutrinos.
The physical effect of $\mathcal{H}_{\text{eff}}$
is described by the effective electromagnetic vertex in Fig.~\ref{A005},
with the neutrino matrix element
\begin{equation}
\langle \nu_{f}(p_{f}) |
j_{\mu}^{\text{eff}}(0)
| \nu_{i}(p_{i}) \rangle
=
\overline{u_{f}}(p_{f})
\Lambda^{fi}_{\mu}(p_{f},p_{i})
u_{i}(p_{i})
.
\label{A027}
\end{equation}
As in the case of one massive neutrino field,
$\Lambda^{fi}_{\mu}(p_{f},p_{i})$
depends only on the four-momentum $q$ transferred to the photon
and can be expressed in terms of six Lorentz-invariant form factors:
\begin{equation}
\Lambda^{fi}_{\mu}(q)
=
f_{1}^{fi}(q^{2}) q_\mu
+
f_{2}^{fi}(q^{2}) q_\mu \gamma_{5}
+
f_{3}^{fi}(q^{2}) \gamma_\mu
+
f_{4}^{fi}(q^{2}) \gamma_\mu \gamma_{5}
+
f_{5}^{fi}(q^{2}) \sigma_{\mu\nu} q^{\nu}
+
f_{6}^{fi}(q^{2}) \epsilon_{\mu\nu\rho\gamma} q^{\nu} \sigma^{\rho\gamma}
.
\label{A028}
\end{equation}
The Hermitian nature of $j_{\mu}^{\text{eff}}$
implies that
$
\langle \nu_{f}(p_{f}) |
j_{\mu}^{\text{eff}}(0)
| \nu_{i}(p_{i}) \rangle
=
\langle \nu_{i}(p_{i}) |
j_{\mu}^{\text{eff}}(0)
| \nu_{f}(p_{f}) \rangle^{*}
$,
leading to the constraint
\begin{equation}
\Lambda^{fi}_{\mu}(q)
=
\gamma^{0} [\Lambda^{if}_{\mu}(-q)]^{\dagger} \gamma^{0}
.
\label{A029}
\end{equation}
Considering the $3\times3$ form factor matrices
$f_{k}$
in the space of massive neutrinos
with components
$f_{k}^{fi}$
for
$k=1,\ldots,6$,
we find that
\begin{equation}
f_{2}
,
\quad
f_{3}
,
\quad
f_{4}
\quad
\text{are Hermitian}
,
\label{A030}
\end{equation}
and
\begin{equation}
f_{1}
,
\quad
f_{5}
,
\quad
f_{6}
\quad
\text{are antihermitian}
.
\label{A031}
\end{equation}

Following the same method used in Eqs.~(\ref{A015})--(\ref{A018}),
one can find that current conservation
implies the constraints
\begin{equation}
f_{1}^{fi}(q^{2}) q^{2}
+
f_{3}^{fi}(q^{2}) (m_{f}-m_{i})
=
0
,
\qquad
f_{2}^{fi}(q^{2}) q^{2}
+
f_{4}^{fi}(q^{2}) (m_{f}+m_{i})
=
0
.
\label{A033}
\end{equation}
Therefore,
we obtain
\begin{equation}
\Lambda^{fi}_{\mu}(q)
=
\left( \gamma_{\mu} - q_{\mu} \sla{q}/q^{2} \right)
\left[
f_{Q}^{fi}(q^{2})
+
f_{A}^{fi}(q^{2}) q^{2} \gamma_{5}
\right]
-
i \sigma_{\mu\nu} q^{\nu}
\left[
f_{M}^{fi}(q^{2})
+ i
f_{E}^{fi}(q^{2}) \gamma_{5}
\right]
,
\label{A035}
\end{equation}
where
$f_{Q}^{fi} = f_{3}^{fi}$,
$f_{M}^{fi} = i f_{5}^{fi}$,
$f_{E}^{fi} = - 2 i f_{6}^{fi}$ and
$f_{A}^{fi} = - f_{2}^{fi} / (m_{f}+m_i)$,
with
\begin{equation}
f_{\Omega}^{fi} = (f_{\Omega}^{if})^{*}
\qquad
(\Omega=Q,M,E,A)
.
\label{A035a}
\end{equation}
Note that since
$
\overline{u_{f}}(p_{f})
\sla{q}
u_{i}(p_{i})
=
(m_{f} - m_{i})
\,
\overline{u_{f}}(p_{f})
u_{i}(p_{i})
$,
if $f=i$
Eq.~(\ref{A035}) correctly reduces to Eq.~(\ref{vert_func}).

The form factors with $f=i$ are called ``diagonal'',
whereas those with $f{\neq}i$ are called ``off-diagonal'' or
``transition form factors''.
This terminology follows from the expression
\begin{equation}
\Lambda_{\mu}(q)
=
\left( \gamma_{\mu} - q_{\mu} \sla{q}/q^{2} \right)
\left[
f_{Q}(q^{2})
+
f_{A}(q^{2}) q^{2} \gamma_{5}
\right]
-
i \sigma_{\mu\nu} q^{\nu}
\left[
f_{M}(q^{2}) 
+ i
f_{E}(q^{2}) \gamma_{5}
\right]
,
\label{A135}
\end{equation}
in which $\Lambda_{\mu}(q)$
is a $3\times3$ matrix
in the space of massive neutrinos expressed
in terms of the four Hermitian $3\times3$ matrices of form factors
\begin{equation}
f_{\Omega} = f_{\Omega}^{\dagger}
\qquad
(\Omega=Q,M,E,A)
.
\label{A134}
\end{equation}

For the coupling with a real photon ($q^{2}=0$) we have
\begin{equation}
f^{fi}_{Q}(0) = \chg_{fi}
,
\
f^{fi}_{M}(0) = \mgm_{fi}
,
\
f^{fi}_{E}(0) = \elm_{fi}
,
\
f^{fi}_{A}(0) = \anm_{fi}
,
\label{A120a}
\end{equation}
where
$\chg_{fi}$,
$\mgm_{fi}$,
$\elm_{fi}$ and
$\anm_{fi}$
are, respectively,
the neutrino charge, magnetic moment, electric moment and anapole moment
of diagonal ($f=i$) and transition ($f{\neq}i$) types.

Considering now CP invariance,
the transformation (\ref{A022a}) of $j_{\mu}^{\text{eff}}(x)$
implies the constraint in Eq.~(\ref{A022})
for the $N{\times}N$ matrix $\Lambda_{\mu}(q)$
in the space of massive neutrinos.
Using Eq.~(\ref{ACP}),
we obtain
\begin{equation}
\Lambda_{\mu}^{fi}(q)
\xrightarrow{\;\text{CP}\;}
\xi^{\text{CP}}_{f} {\xi^{\text{CP}}_{i}}^{*}
\gamma^{0}
\mathcal{C}
[\Lambda_{\mu}^{if}(q_{\text{P}})]^{T}
\mathcal{C}^{\dagger}
\gamma^{0}
,
\label{A123a}
\end{equation}
where
$\xi^{\text{CP}}_{k}$
is the CP phase of $\nu_{k}$.
Since the three massive neutrinos take part to standard charged-current weak
interactions,
their CP phases are equal if CP is conserved
(see \cite{Giunti-Kim-2007}).
Hence, we have
\begin{equation}
\Lambda_{\mu}^{fi}(q)
\xrightarrow{\;\text{CP}\;}
\gamma^{0}
\mathcal{C}
[\Lambda_{\mu}^{if}(q_{\text{P}})]^{T}
\mathcal{C}^{\dagger}
\gamma^{0}
.
\label{A123}
\end{equation}
Using the form factor expansion in Eq.~(\ref{A035}),
we obtain
\begin{equation}
\Lambda_{\mu}^{fi}(q)
\xrightarrow{\;\text{CP}\;}
-
\Big\{
\left( \gamma^{\mu} - q^{\mu} \sla{q}/q^{2} \right)
\left[
f_{Q}^{if}(q^{2})
+
f_{A}^{if}(q^{2}) q^{2} \gamma_{5}
\right]
-
i \sigma^{\mu\nu} q_{\nu}
\left[
f_{M}^{if}(q^{2})
- i
f_{E}^{if}(q^{2}) \gamma_{5}
\right]
\Big\}
.
\label{A124}
\end{equation}
Imposing the constraint in Eq.~(\ref{A022}),
for the form factors we obtain
\begin{equation}
\text{CP}
\quad
\Longleftrightarrow
\quad
\left\{
\begin{array}{l} \displaystyle
f_{\Omega}^{fi} = f_{\Omega}^{if} = (f_{\Omega}^{fi})^{*}
\qquad
(\Omega=Q,M,A)
,
\\ \displaystyle
f_{E}^{fi} = - f_{E}^{if} = - (f_{E}^{fi})^{*}
,
\end{array}
\right.
\label{A036}
\end{equation}
where, in the last equalities, we took into account the constraints (\ref{A035a}).
For the Hermitian $3\times3$ form factor matrices we obtain that
if CP is conserved
$f_{Q}$,
$f_{M}$ and
$f_{A}$
are real and symmetric
and
$f_{E}$ is imaginary and antisymmetric:
\begin{equation}
\text{CP}
\quad
\Longleftrightarrow
\quad
\left\{
\begin{array}{l} \displaystyle
f_{\Omega} = f_{\Omega}^{\text{T}} = f_{\Omega}^{*}
\qquad
(\Omega=Q,M,A)
,
\\ \displaystyle
f_{E} = - f_{E}^{\text{T}} = - f_{E}^{*}
.
\end{array}
\right.
\label{A037}
\end{equation}

Let us now consider antineutrinos.
Using for the massive neutrino fields the Fourier expansion in Eq.~(2.139) of \cite{Giunti-Kim-2007},
the effective antineutrino matrix element
for
$\bar\nu_{i}(p_{i}) \to \bar\nu_{f}(p_{f})$
transitions is given by
\begin{equation}
\langle \bar\nu_{f}(p_{f}) |
j_{\mu}^{\text{eff}}(0)
| \bar\nu_{i}(p_{i}) \rangle
=
-
\overline{v_{i}}(p_{i})
\Lambda^{if}_{\mu}(q)
v_{f}(p_{f})
.
\label{A327}
\end{equation}
Using the relation
\begin{equation}
u(p)
=
\mathcal{C} \, \overline{v}^{T}(p)
,
\label{AB032}
\end{equation}
we can write it as
\begin{equation}
\langle \bar\nu_{f}(p_{f}) |
j_{\mu}^{\text{eff}}(0)
| \bar\nu_{i}(p_{i}) \rangle
=
\overline{u_{f}}(p_{f})
\mathcal{C}
[\Lambda^{if}_{\mu}(q)]^{T}
\mathcal{C}^{\dagger}
u_{i}(p_{i})
,
\label{A328}
\end{equation}
where transposition operates in spinor space.
Therefore,
the effective form factor matrix in spinor space for antineutrinos is given by
\begin{equation}
\overline{\Lambda}^{fi}_{\mu}(q)
=
\mathcal{C}
[\Lambda^{if}_{\mu}(q)]^{T}
\mathcal{C}^{\dagger}
.
\label{A329}
\end{equation}
Using the properties of the charge-conjugation matrix and the expression (\ref{A035}) for $\Lambda^{if}_{\mu}(q)$,
we obtain
the antineutrino form factors
\begin{align}
\null & \null
\overline{f}_{\Omega}^{fi} = - f_{\Omega}^{if}
\qquad
(\Omega=Q,M,E)
,
\label{A335a}
\\
\null & \null
\overline{f}_{A}^{fi} = f_{A}^{if}
.
\label{A335b}
\end{align}
Therefore,
in particular
the diagonal magnetic and electric moments of
neutrinos and antineutrinos have the same size with opposite signs,
as the charge, if it exists.
On the other hand,
the diagonal neutrino and antineutrino anapole moments are equal.

\subsection{Majorana neutrinos}
\label{ff:Majorana}

A massive Majorana neutrino is a neutral spin 1/2 particle which coincides with its antiparticle.
The four degrees of freedom of a massive Dirac field
(two helicities and two particle-antiparticle)
are reduced to two
(two helicities) by the Majorana constraint
\begin{equation}
\nu_{k}
=
\nu_{k}^{c}
=
\mathcal{C} \overline{\nu_{k}}^{T}
.
\label{A039}
\end{equation}
Since a Majorana field has half the degrees of freedom of a Dirac field,
it is possible that its electromagnetic properties are reduced.
From the relations (\ref{A335a}) and (\ref{A335b})
between neutrino and antineutrino form factors in the Dirac case,
we can infer that in the Majorana case
the charge, magnetic and electric form factor matrices are antisymmetric and
the anapole form factor matrix is symmetric.
In order to confirm this deduction, let us calculate the neutrino matrix element
corresponding to
the effective electromagnetic vertex in Fig.~\ref{A005},
with the effective interaction Hamiltonian in Eq.~(\ref{A026}),
which takes into account possible transitions
between two different initial and final massive Majorana neutrinos $\nu_{i}$ and $\nu_{f}$.
Using
for the neutrino Majorana fields
the Fourier expansion in Eq.~(6.99) of \cite{Giunti-Kim-2007}, we obtain
\begin{equation}
\langle \nu_{f}(p_{f}) |
j_{\mu}^{\text{eff}}(0)
| \nu_{i}(p_{i}) \rangle
=
\overline{u_{f}}(p_{f})
\Lambda^{fi}_{\mu}(p_{f},p_{i})
u_{i}(p_{i})
-
\overline{v_{i}}(p_{i})
\Lambda^{if}_{\mu}(p_{f},p_{i})
v_{f}(p_{f})
.
\label{A040}
\end{equation}
Using Eq.~(\ref{AB032}),
we can write it as
\begin{equation}
\overline{u_{f}}(p_{f})
\left\{
\Lambda^{fi}_{\mu}(p_{f},p_{i})
+
\mathcal{C}
[\Lambda^{if}_{\mu}(p_{f},p_{i})]^{T}
\mathcal{C}^{\dagger}
\right\}
u_{i}(p_{i})
,
\label{A041}
\end{equation}
where transposition operates in spinor space.
Therefore the effective
form factor matrix in spinor space for Majorana neutrinos is given by
\begin{equation}
\widetilde{\Lambda}^{fi}_{\mu}(p_{f},p_{i})
=
\Lambda^{fi}_{\mu}(p_{f},p_{i})
+
\mathcal{C}
[\Lambda^{if}_{\mu}(p_{f},p_{i})]^{T}
\mathcal{C}^{\dagger}
.
\label{A042}
\end{equation}
As in the case of Dirac neutrinos,
$\Lambda^{fi}_{\mu}(p_{f},p_{i})$
depends only on $q=p_{f}-p_{i}$
and can be expressed in terms of six Lorentz-invariant form factors
according to Eq.~(\ref{A028}).
Hence, we can write
the $3\times3$ matrix $\widetilde{\Lambda}_{\mu}(p_{f},p_{i})$
in the space of massive Majorana neutrinos
as
\begin{equation}
\widetilde{\Lambda}_{\mu}(q)
=
\widetilde{f}_{1}(q^{2}) q_\mu
+
\widetilde{f}_{2}(q^{2}) q_\mu \gamma_{5}
+
\widetilde{f}_{3}(q^{2}) \gamma_\mu
+
\widetilde{f}_{4}(q^{2}) \gamma_\mu \gamma_{5}
+
\widetilde{f}_{5}(q^{2}) \sigma_{\mu\nu} q^{\nu}
+
\widetilde{f}_{6}(q^{2}) \epsilon_{\mu\nu\rho\gamma} q^{\nu} \sigma^{\rho\gamma}
,
\label{A128}
\end{equation}
with
\begin{align}
\null & \null
\widetilde{f}_{k} = f_{k} + f_{k}^{T}
\Longrightarrow
\widetilde{f}_{k} = \widetilde{f}_{k}^{T}
\null & \null
\text{for}
\quad
k=1,2,4
,
\label{A128a}
\\
\null & \null
\widetilde{f}_{k} = f_{k} - f_{k}^{T}
\Longrightarrow
\widetilde{f}_{k} = - \widetilde{f}_{k}^{T}
\null & \null
\text{for}
\quad
k=3,5,6
.
\label{A128b}
\end{align}
Now we can follow the discussion in Section~\ref{ff:Dirac}
for Dirac neutrinos taking into account the additional constraints
(\ref{A128a}) and (\ref{A128b})
for Majorana neutrinos.
The hermiticity of $j_{\mu}^{\text{eff}}$
and current conservation lead to an expression similar to that
in Eq.~(\ref{A135}):
\begin{equation}
\widetilde{\Lambda}_{\mu}(q)
=
\left( \gamma_{\mu} - q_{\mu} \sla{q}/q^{2} \right)
\left[
\widetilde{f}_{Q}(q^{2})
+
\widetilde{f}_{A}(q^{2}) q^{2} \gamma_{5}
\right]
-
i \sigma_{\mu\nu} q^{\nu}
\left[
\widetilde{f}_{M}(q^{2})
+ i
\widetilde{f}_{E}(q^{2}) \gamma_{5}
\right]
,
\label{A043}
\end{equation}
with
$\widetilde{f}_{Q}^{fi} = \widetilde{f}_{3}^{fi}$,
$\widetilde{f}_{M}^{fi} = i \widetilde{f}_{5}^{fi}$,
$\widetilde{f}_{E}^{fi} = - 2 i \widetilde{f}_{6}^{fi}$ and
$\widetilde{f}_{A}^{fi} = - \widetilde{f}_{2}^{fi} / (m_{f}+m_i)$.
For the Hermitian $3\times3$ form factor matrices
in the space of massive neutrinos,
\begin{equation}
\widetilde{f}_{\Omega} = \widetilde{f}_{\Omega}^{\dagger}
\qquad
(\Omega=Q,M,E,A)
,
\label{A234}
\end{equation}
the Majorana constraints (\ref{A128a}) and (\ref{A128b})
imply that
\begin{align}
\null & \null
\widetilde{f}_{\Omega} = - \widetilde{f}_{\Omega}^{T}
\quad
(\Omega=Q,M,E)
,
\label{A044}
\\
\null & \null
\widetilde{f}_{A} = \widetilde{f}_{A}^{T}
.
\label{A045}
\end{align}
These relations
confirm the expectation discussed above that
for Majorana neutrinos
the charge, magnetic and electric form factor matrices are antisymmetric and
the anapole form factor matrix is symmetric.

Since
$\widetilde{f}_{Q}$,
$\widetilde{f}_{M}$ and
$\widetilde{f}_{E}$
are antisymmetric, a Majorana neutrino does not have
diagonal charge and dipole magnetic and electric form factors.
It can only have a diagonal anapole form factor.
On the other hand,
Majorana neutrinos can have as many off-diagonal (transition) form factors as Dirac neutrinos.

Since the form factor matrices are Hermitian as in the Dirac case,
$\widetilde{f}_{Q}$,
$\widetilde{f}_{M}$ and
$\widetilde{f}_{E}$
are imaginary,
whereas
$\widetilde{f}_{A}$ is real:
\begin{align}
\null & \null
\widetilde{f}_{\Omega} = - \widetilde{f}_{\Omega}^{*}
\qquad
(\Omega=Q,M,E)
,
\label{A046}
\\
\null & \null
\widetilde{f}_{A} = \widetilde{f}_{A}^{*}
.
\label{A047}
\end{align}

Considering now CP invariance,
the case of Majorana neutrinos is rather different from that of Dirac neutrinos,
because the CP phases of the massive Majorana fields
$\nu_{k}$
are constrained by the CP invariance of the Lagrangian Majorana mass term
\begin{equation}
\mathscr{L}_{\text{M}}
=
\frac{1}{2} \sum_{k}
m_{k} \, \nu_{k}^{T} \, \mathcal{C}^{\dagger} \, \nu_{k}
.
\label{022}
\end{equation}
In order to prove this statement,
let us first notice that since a massive Majorana neutrino field
$\nu_{k}$
is constrained by the Majorana relation in Eq.~(\ref{A039}),
only the parity transformation part is effective in a CP transformation:
\begin{equation}
\mathsf{U}_{\text{CP}}
\nu_{k}(x)
\mathsf{U}_{\text{CP}}^{\dagger}
=
\xi^{\text{CP}}_{k} \gamma^{0} \nu_{k}(x_{\text{P}})
.
\label{A048a}
\end{equation}
Considering the mass term in Eq.~(\ref{022}),
we have
\begin{equation}
\mathsf{U}_{\text{CP}}
\nu_{k}^{T} \, \mathcal{C}^{\dagger} \, \nu_{k}
\mathsf{U}_{\text{CP}}^{\dagger}
=
- {\xi^{\text{CP}}_{k}}^{2}
\,
\nu_{k}^{T} \, \mathcal{C}^{\dagger} \, \nu_{k}
.
\label{A048}
\end{equation}
Therefore,
\begin{equation}
\text{CP}
\quad
\Longleftrightarrow
\quad
\xi^{\text{CP}}_{k}
=
\eta_{k} \, i
,
\label{A049}
\end{equation}
with
$\eta_{k} = \pm 1$.
These CP signs can be different for the different massive neutrinos,
even if they all take part to the standard charged-current weak interactions
through neutrino mixing,
because they can be compensated by the Majorana CP phases in the mixing matrix
(see \cite{Giunti-Kim-2007}).
Therefore,
from Eq.~(\ref{A123a}) we have
\begin{equation}
\widetilde{\Lambda}_{\mu}^{fi}(q)
\xrightarrow{\;\text{CP}\;}
\eta_{f} \eta_{i}
\gamma^{0}
\mathcal{C}
[\widetilde{\Lambda}_{\mu}^{if}(q_{\text{P}})]^{T}
\mathcal{C}^{\dagger}
\gamma^{0}
.
\label{A223a}
\end{equation}
Imposing a CP constraint analogous to that in Eq.~(\ref{A022}),
we obtain
\begin{equation}
\text{CP}
\quad
\Longleftrightarrow
\quad
\left\{
\begin{array}{l} \displaystyle
f_{\Omega}^{fi} = \eta_{f} \eta_{i} f_{\Omega}^{if} = \eta_{f} \eta_{i} (f_{\Omega}^{fi})^{*}
,
\\ \displaystyle
f_{E}^{fi} = - \eta_{f} \eta_{i} f_{E}^{if} = - \eta_{f} \eta_{i} (f_{E}^{fi})^{*}
,
\end{array}
\right.
\label{A236}
\end{equation}
with
$\Omega=Q,M,A$.
Taking into account the constraints (\ref{A046}) and (\ref{A047}),
we have two cases:
\begin{equation}
\text{CP}
\quad
\text{and}
\quad
\eta_{f} = \eta_{i}
\quad
\Longleftrightarrow
\quad
f_{Q}^{fi} = f_{M}^{fi} = 0
,
\label{A050}
\end{equation}
and
\begin{equation}
\text{CP}
\quad
\text{and}
\quad
\eta_{f} = - \eta_{i}
\quad
\Longleftrightarrow
\quad
f_{E}^{fi} = f_{A}^{fi} = 0
.
\label{A051}
\end{equation}
Therefore,
if CP is conserved
two massive Majorana neutrinos can have either
a transition electric form factor
or
a transition magnetic form factor,
but not both,
and
the transition electric form factor can exist only together with a transition anapole form factor,
whereas
the transition magnetic form factor can exist only together with a transition charge form factor.
In the diagonal case $f=i$,
Eq.~(\ref{A050}) does not give any constraint,
because only diagonal
anapole form factors are allowed for Majorana neutrinos.

\subsection{Form factors in gauge models}
\label{form_fac_gauge_mod}

From the demand that the form factors at
zero momentum transfer, $q^{2}=0$, are elements of the scattering
matrix, it follows that in any consistent theoretical model the
form factors in the matrix element (\ref{matr_elem}) should be
gauge independent and finite. Then, the form factors values at
$q^{2}=0$ determine the static electromagnetic properties of the
neutrino that can be probed or measured in the direct interaction
with external electromagnetic fields. This is the case for charge,
dipole magnetic and electric neutrino form factors in the
minimally extended Standard Model.

In non-Abelian gauge theories, the form factors in the matrix
element (\ref{matr_elem}) at nonzero momentum transfer, $q^{2}\neq
0$, can be non-invariant under gauge transformations. This
happens because in general the off-shell photon propagator is
gauge dependent. Therefore, the one-photon approximation is not
enough to get physical quantities. In this case the form factors
in the matrix element (\ref{matr_elem}) cannot be directly
measured in an experiment with an external electromagnetic field.
However, they can contribute to higher-order diagrams describing some
processes that are accessible for experimental observation
(see \cite{Bardeen:1972vi}).

Note that there is an important difference between
the electromagnetic vertex function of massive and massless
neutrinos \cite{Dvornikov:2003js,Dvornikov:2004sj}. For the case
of a massless neutrino, the matrix element (\ref{matr_elem}) of the electromagnetic
current can be expressed in terms of only one
Dirac form factor $f_D (q^{2})$ (see also \cite{NovalesSanchez:2008tn}),
\begin{equation}
{\bar u}(p^{\prime})\Lambda_{\mu}(q)u(p)=
f_{D}(q^{2}){\bar u}(p^{\prime})\gamma_{\mu}(1+\gamma_{5})u(p).
\end{equation}
It follows that the electric charge and anapole form factors for a massless
neutrino are related to the Dirac form factor $f_{D}(q^{2})$, and hence to each
other:
\begin{equation}
f_{Q}(q^{2})=f_{D}(q^{2}), \quad f_{A}(q^{2})=f_{D}(q^{2})/q^{2}.
\end{equation}

In the case of a massive neutrino, there is no such simple relation between
electric charge and anapole form factors since the
$q_{\mu}\sla{q}\gamma_{5}$ term in the anapole part of the vertex function
(\ref{vert_func}) cannot be neglected.

Moreover,
a direct calculation
of the massive neutrino electromagnetic vertex function, taking
into account all the diagrams
in Figs.~(15)--(17) of \cite{Giunti:2008ve},
%(Fig.~\ref{fig1f} and Fig.~\ref{fig2}),
reveals that each of the Feynman diagrams gives
nonzero contribution to the term proportional to
$\gamma_{\mu}\gamma_{5}$ \cite{Dvornikov:2003js,Dvornikov:2004sj}.
These contributions are not vanishing
even at $q^{2}=0$. Therefore, in addition to the usual four terms
in (\ref{vert_func}) an extra term proportional to
$\gamma_{\mu}\gamma_{5}$ appears and a corresponding additional
form factor $f_5 (q^{2})$ must be introduced. This problem is related
to the decomposition of the massive neutrino electromagnetic
vertex function.
The calculation of the contributions of the
proper vertex diagrams
(Fig.~(15) of \cite{Giunti:2008ve})
%(Fig.~\ref{fig1f})
and $\gamma-Z$
self-energy diagrams
(Fig.~(16) and (17) of \cite{Giunti:2008ve})
%(Fig.~\ref{fig4_0})
for arbitrary gauge
fixing parameter $\alpha=1/\xi$
in the
general $R_{\xi}$ gauge and arbitrary mass
parameter $a=m_l^{2}/m_{W}^{2}$ shows that at least in the
zeroth and first orders of the expansion over the small neutrino
mass parameter $b=(m_{\nu}/m_{W})^{2}$ the corresponding
``charge'' $f_5 (q^{2}=0)$ is zero. The cancellation of
contributions from the proper vertex and self-energy diagrams to
the form factor $f_5 (q^{2})$ at $q^{2}\neq0$,
\begin{equation}
f_{5}(q^{2})=f_{5}^{(\gamma-Z)}(q^{2})+
f_{5}^{(\mathrm{prop. vert.})}(q^{2})=0,
\end{equation}
was also shown \cite{Dvornikov:2003js,Dvornikov:2004sj} for
arbitrary mass parameters $a$ and $b$ in the `t Hooft-Feynman
gauge $\alpha=1$.

Hence,
in the
minimally extended Standard Model
one can perform a direct calculation of the
neutrino vertex function leading to the four terms
in (\ref{vert_func})
with gauge-invariant electric charge, magnetic, electric and anapole moments.
\section{Magnetic and electric dipole moments}
\label{sec4}

The neutrino dipole magnetic and electric form factors (and the
corresponding magnetic and electric dipole moments) are theoretically
the most well-studied and understood among the form factors. They
also attract a reasonable attention from experimentalists, although
the neutrino magnetic moment predicted in the
extended Standard Model with right-handed neutrinos is
proportional to the neutrino mass and therefore it is many orders of
magnitude smaller than the present experimental limits obtained in
terrestrial experiments.

\subsection{Theoretical predictions}
\label{sec:theoretical}

The first calculations of the
neutrino dipole moments within the minimal extension of the
Standard Model with right-handed neutrinos
were performed in
\cite{Marciano:1977wx,Lee:1977tib,Fujikawa:1980yx,Petcov:1976ff,Pal:1981rm,Shrock:1982sc}.
The explicit evaluation of the one-loop
contributions to the neutrino dipole moments in the leading
approximation over the small parameters
$b_i = m_{i}^{2} /m_{W}^{2}$ (where $m_i$ are the neutrino
masses, $i=1,2,3$), that in addition exactly accounts for the
dependence on the small parameters
$a_l = m_{l}^{2} / m_{W}^{2}$ (with $l=e,\mu,\tau$), yields, for
Dirac neutrinos
\cite{Petcov:1976ff,Pal:1981rm,Shrock:1982sc,Bilenky:1987ty,Mohapatra:2004},
\begin{equation}\label{m_e_mom_i_j}
\left.
\begin{array}{c}
\mu^{\text{D}}_{ij}
\\
\epsilon^{\text{D}}_{ij}
\end{array}
\right\}
=
\frac{e G_F}{8\sqrt{2}\pi^{2}}
\left( m_{i} \pm m_{j} \right)
\sum_{l=e,\mu,\tau} f(a_l) U^{*}_{li} U_{lj}
,
\end{equation}
where
\begin{equation}\label{f_a_l}
f(a_l)
=
\frac{3}{4}
\left[1+\frac{1}{1-a_l}-\frac{2a_l}{(1-a_l)^{2}}-\frac{2a_l^{2}\ln a_l}{(1-a_l)^3}\right]
.
\end{equation}
All the charged lepton parameters $a_l$ are small. In the limit
$a_l\ll 1$, one has
\begin{equation}\label{f_appr}
f(a_l)
\simeq
\frac{3}{2}
\left(1 - \frac{a_l}{2}\right)
.
\end{equation}
From Eqs.~(\ref{m_e_mom_i_j}) and (\ref{f_appr}), the diagonal
magnetic moments of Dirac neutrinos are given by
\begin{equation}\label{nu_mu_D_ii}
\mu^{\text{D}}_{ii}
\simeq
\frac{3e G_F m_{i}}{8\sqrt{2} \pi^{2}}
\left(
1
-
\frac{1}{2} \sum_{l=e,\mu,\tau}
a_l
|U_{li}|^{2}
\right)
.
\end{equation}
This result exhibits the following important features.
The magnetic moment of a Dirac neutrino is proportional to the neutrino
mass and for a massless Dirac neutrino in the Standard Model (in the
absence of right-handed charged currents) the magnetic moment is
zero. The magnetic moment of a massive Dirac neutrino, at the leading
order in $a_l$, is independent of the neutrino mixing matrix and
of the values of the charged lepton masses. The numerical
value of the Dirac neutrino magnetic moment is
\begin{equation}\label{mu_3_10_19}
\mu^{\text{D}}_{ii}
\simeq
3.2 \times 10^{-19}
\left( \frac{m_i}{\text{eV}} \right) \mu_{B}
.
\end{equation}
Taking into account the existing constraint of the order of 1 eV on the neutrino masses
(see
\cite{Giunti-Kim-2007,GonzalezGarcia:2007ib,Bilenky:2010zza,Xing:2011zza}),
this value is several orders of magnitude smaller than the present experimental limits,
which are discussed in Section~\ref{sec:limits}.

From Eq.~(\ref{m_e_mom_i_j}), it can be clearly seen that in the
extended Standard Model with right-handed neutrinos
the static (diagonal) electric dipole moment of a
Dirac neutrino vanishes, $\epsilon^{\text{D}}_{ii}= 0$,
in spite of possible CP violations generated by the Dirac phase in the mixing matrix
(as shown in Eq.~(\ref{A025}),
Dirac neutrinos may
have nonzero diagonal electric moments only in theories where CP invariance is violated).
For a Majorana neutrino both the diagonal
magnetic and electric moments are zero,
$\mu^{\text{M}}_{ii}=\epsilon^{\text{M}}_{ii}=0$,
as shown in Section~\ref{ff:Majorana}.

Let us consider now the neutrino transition moments, which are
given by Eq.~(\ref{m_e_mom_i_j}) for $i\neq j$.
Considering only
the leading term
$f(a_l) \simeq 3/2$
in the expansion (\ref{f_appr}),
one gets vanishing transition moments,
because of the unitarity relation
\begin{equation}
\sum_{l} U^{*}_{li} U_{lj}
=
\delta_{ij}
.
\label{unitarity}
\end{equation}
Therefore,
the first nonvanishing contribution comes from the second term in the expansion
(\ref{f_appr})
of $f(a_l)$, which contains the additional small factor
$a_l=m_{l}^{2}/m_{W}^{2}$:
\begin{equation}\label{m_e_mom_i_not_j}
\left.
\begin{array}{c}
\mu^{\text{D}}_{ij}
\\
\epsilon^{\text{D}}_{ij}
\end{array}
\right\}
\simeq
-
\frac{3e G_F}{32\sqrt{2}\pi^{2}}
\left( m_{i} \pm m_{j} \right)
\sum_{l=e,\mu,\tau}
\left(\frac{m_l}{m_{W}}\right)^{2}
U^{*}_{li} U_{lj}
,
\end{equation}
for $i \neq j$.
Thus, the transition moments are suppressed with respect to the diagonal
magnetic moments in Eq.~(\ref{nu_mu_D_ii}).
This suppression is called
``GIM mechanism'',
in analogy with the suppression of flavor-changing neutral currents
in hadronic processes discovered in
\cite{Glashow:1970gm}.
Numerically, the Dirac transition moments
are given by
\begin{equation}
\left.
\begin{array}{c}
\mu^{\text{D}}_{ij}
\\
\epsilon^{\text{D}}_{ij}
\end{array}
\right\}
\simeq
-
4 \times 10^{-23}
\left(\frac{m_i \pm m_j}{\text{eV}}\right)
f_{ij} \, \mu_{B}
,
\end{equation}
with
\begin{equation}
f_{ij}
=
\sum_{l=e,\mu,\tau}
\left(\frac{m_l}{m_\tau}\right)^{2} U^{*}_{li} U_{lj}
.
\label{fij}
\end{equation}

Also Majorana neutrinos can have nonvanishing transition magnetic
and electric moments, as discussed in
Section~\ref{ff:Majorana}.
Assuming CP conservation
and neglecting model-dependent Feynman diagrams depending on the details of the scalar sector
\cite{Schechter:1981hw,Pal:1981rm,Shrock:1982sc,Mohapatra:2004},
if $\nu_{i}$ and $\nu_{j}$
have the same CP phase,
\begin{equation}
\mu^{\text{M}}_{ij} = 0
\qquad
\text{and}
\qquad
\epsilon^{\text{M}}_{ij}
=
2 \epsilon^{\text{D}}_{ij}
,
\label{edm:Majorana}
\end{equation}
whereas
if $\nu_{i}$ and $\nu_{j}$
have opposite CP phases,
\begin{equation}
\mu^{\text{M}}_{ij}
=
2 \mu^{\text{D}}_{ij}
\qquad
\text{and}
\qquad
\epsilon^{\text{D}}_{ij} = 0
,
\label{mdm:Majorana}
\end{equation}
with
$\epsilon^{\text{D}}_{ij}$
and
$\mu^{\text{D}}_{ij}$
given by Eq.~(\ref{m_e_mom_i_j}).
Hence,
although the non-vanishing Majorana transition moments are twice the Dirac ones,
they are equally suppressed by the GIM mechanism.
However,
the model-dependent contributions of the scalar sector
can enhance the Majorana transition moments
(see \cite{Pal:1981rm,Barr:1990um,Pal:1991qr}).

In recent studies, the value of the diagonal
magnetic moment of a massive Dirac neutrino
was calculated in the one-loop approximation in the
extended Standard Model with right-handed neutrinos,
accounting for the dependence on the neutrino
mass parameter $b_i=m_{i}^{2}/m_{W}^{2}$
\cite{CabralRosetti:1999ad} and accounting for the exact
dependence on both mass parameters $b_i$ and
$a_l=m_{l}^{2}/m_{W}^{2}$
\cite{Dvornikov:2003js,Dvornikov:2004sj}. The calculations of the neutrino magnetic
moment which take into account exactly the dependence on the
masses of all particles can be useful in the case of a heavy
neutrino with a mass comparable or even exceeding the values of
the masses of other known particles.
Note that the LEP data require that
the number of light neutrinos coupled to the $Z$ boson is three
\cite{hep-ex/0509008}. Therefore, any additional active neutrino must be heavier
than $m_{Z}/2$. This possibility is not
excluded by current data
(see \cite{1112.2907}).

For a heavy neutrino with mass $m_{i}$
much larger than the charged lepton masses
but smaller than the $W$-boson mass
($2 \, \text{GeV} \ll m_{i} \ll 80 \, \text{GeV}$),
the authors of \cite{Dvornikov:2003js,Dvornikov:2004sj} obtained
the diagonal magnetic moment
\begin{equation}
\mu_{ii}
\simeq
\frac{3eG_{F}}{8\pi^{2}\sqrt{2}}
\,
m_{i}
\left(
1+{\frac{5}{18}}b_{i}
\right)
,
\label{DS-04-1}
\end{equation}
whereas for a heavy neutrino with mass $m_{i}$
much larger than the $W$-boson mass, they got
\begin{equation}
\mu_{ii}
\simeq
\frac{eG_{F}}{8\pi^{2}\sqrt{2}}
\,
m_{i}
.
\label{DS-04-2}
\end{equation}
Note that in both cases the Dirac neutrino magnetic
moment is proportional to the neutrino mass. This is an expected
result, because the calculations have been performed within the
extended Standard Model with right-handed neutrinos.

At this point, a question arises: ``Is a neutrino magnetic
moment always proportional to the neutrino mass?''. The answer is
``No''. For example, much larger values of the Dirac neutrino
magnetic moment can be obtained in $SU(2)_L \times SU(2)_R \times U(1)$
left-right symmetric models with
direct right-handed neutrino interactions
(see \cite{Kim:1976gk,Marciano:1977wx,Czakon:1998rf,Beg:1977xz}).
The massive gauge bosons states $W_1$ and $W_2$ have, respectively, predominant
left-handed and right-handed coupling, since
\begin{equation}
W_1=W_{L}\cos \xi - W_R \sin \xi
,
\qquad
W_2=W_{L}\sin \xi + W_R \cos \xi
,
\label{left-right}
\end{equation}
where $\xi$ is a small mixing angle and the fields $W_L$ and $W_R$ have pure
$V \pm A$ interactions. The magnetic moment of a
neutrino $\nu_l$ calculated in this model, neglecting neutrino mixing, is
\begin{equation}
\mu_{\nu_{l}}
=
\frac{eG_F}{2\sqrt{2}\pi^{2}}
\left[
m_l
\left(1-\frac{m_{W_1}^{2}}{m_{W_2}^{2}}\right)
\sin2\xi
+
\frac{3}{4}
m_{\nu_{l}}
\left(1+\frac{m_{W_1}^{2}}{m_{W_2}^{2}}\right)
\right]
.
\label{mu_L_R}
\end{equation}
where the term proportional to the charged lepton mass $m_l$ is
due to the left-right mixing. This term can exceed the second term
in Eq.~(\ref{mu_L_R}), which is proportional to the neutrino mass
$m_{\nu_{l}}$.

\subsection{Neutrino-electron elastic scattering}
\label{sec3.6}

The most sensitive and widely used method for the experimental
investigation of the neutrino magnetic moment is provided by
direct laboratory measurements of low-energy elastic scattering of neutrinos and antineutrinos
with electrons in
reactor, accelerator and solar experiments.
Detailed descriptions
of several experiments can be found in
\cite{Wong:2005pa,Beda:2007hf}.

Extensive experimental studies of the neutrino magnetic moment,
performed during many years, are stimulated by the hope to observe
a value much larger than the prediction in Eq.~(\ref{mu_3_10_19}) of the
minimally extended Standard Model with right-handed neutrinos.
It would be a clear indication of new physics beyond the extended
Standard Model.
For example,
the effective magnetic moment in
$\bar\nu_{e}$-$e$ elastic scattering
in a class of extra-dimension models
can be as large as
$\sim 10^{-10} \mu_{B}$
\cite{Mohapatra:2004ce}.
Future higher precision
reactor experiments can therefore be used to provide new
constraints on large extra-dimensions.

The possibility for neutrino-electron elastic scattering due to neutrino
magnetic moment was first considered in \cite{Carlson:1932rk}
and the cross section of this process was calculated in
\cite{Bethe:1935cp,Domogatsky:1971tu}.
Discussions on the derivation of the cross section
and on the optimal conditions for bounding the neutrino magnetic
moment, as well as a collection of cross section formulas for
elastic scattering of neutrinos (antineutrinos) on electrons, nucleons,
and nuclei can be found in \cite{Kyuldjiev:1984kz,Vogel:1989iv}.

Let us consider the elastic scattering
\begin{equation}
\nu + e^{-} \to \nu + e^{-}
\label{elastic}
\end{equation}
of a neutrino with energy $E_{\nu}$
with an electron at rest in the laboratory frame.
There are two observables:
the kinetic
energy $T$
of the recoil electron
and the recoil angle $\chi$
with respect to the neutrino beam,
which are related by
\begin{equation}\cos \chi = \frac {E_{\nu}+m_{e}}{E_{\nu}}\Big
[\frac {T}{T+2m_{e}}\Big]^{1/2}
.
\end{equation}
The electron kinetic energy is constrained from the energy-momentum
conservation by
\begin{equation}
T \leq \frac {2E_{\nu}^{2}}{2E_{\nu} + m_{e}}
.
\end{equation}

Since,
in the ultrarelativistic limit,
the neutrino
magnetic moment interaction changes the neutrino helicity
and the Standard Model weak interaction conserves the neutrino helicity,
the two contributions add incoherently in the cross section
which can be written as
\cite{Vogel:1989iv},
\begin{equation}\label{d_sigma}
\frac{d\sigma}{dT}=\left(\frac{d\sigma}{dT}\right)_{\text{SM}}+
\left(\frac{d\sigma}{dT}\right)_{\mu}
.
\end{equation}
The small interference term
due to neutrino masses has been derived in \cite{Grimus:1997aa}.

The weak-interaction cross section is given by
\begin{equation}
\left(\frac{d\sigma}{dT}\right)_{\text{SM}}
=
\frac{G^{2}_F m_{e}}{2\pi}
\bigg[
(g_V + g_A)^{2}
+
(g_V - g_A)^{2}
\left(1-\frac{T}{E_{\nu}}\right)^{2}
+
(g_A^{2} - g_V^{2})
\frac{m_{e}T}{E^{2}_{\nu}}
\bigg]
,
\label{d_sigma_SM}
\end{equation}
with the standard coupling constants $g_V$ and $g_A$ given by
\begin{align}
\null & \null
g_V
=
\left\{
\begin{array}{ll} \displaystyle
2\sin^{2} \theta_{W} +1/2 & \quad \text{for} \quad \nu_{e},
\\
2\sin^{2} \theta_{W} -1/2 & \quad \text{for} \quad \nu_\mu,\nu_\tau,
\end{array}
\right.
\label{gV}
\\
\null & \null
g_A
=
\left\{
\begin{array}{ll} \displaystyle
1/2 & \quad \text{for} \quad \nu_{e},
\\
-1/2 & \quad \text{for} \quad \nu_\mu,\nu_\tau.
\end{array}
\right.
\label{gA}
\end{align}
In antineutrino-electron elastic scattering,
one must substitute
$g_A \to -g_A$.

The neutrino magnetic-moment contribution to the cross section is given by
\cite{Vogel:1989iv}
\begin{equation}\label{d_sigma_mu}
\left(\frac{d\sigma}{dT}\right)_{\mu}
=
\frac{\pi\alpha^{2}}{m_{e}^{2}}
\left(\frac{1}{T}-\frac{1}{E_{\nu}}\right)
\left(\frac{\mu_{\nu}}{\mu_{B}}\right)^{2}
,
\end{equation}
where $\mu_{\nu}$ is the effective magnetic moment discussed in Section~\ref{sec:effective}.

\begin{figure}
\begin{center}
\includegraphics*[width=0.5\linewidth]{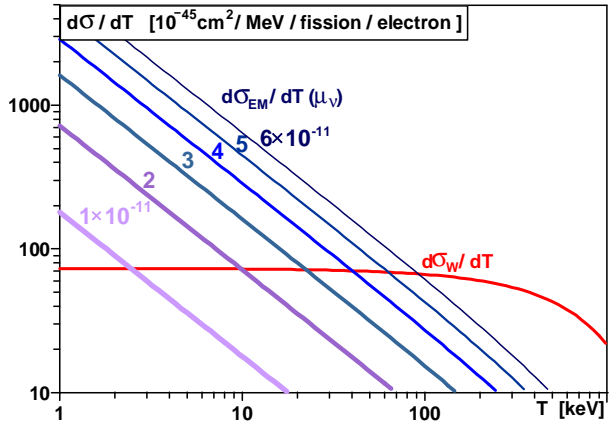}
\end{center}
\caption{\label{gemma_f1}
Standard Model weak (W) and magnetic moment electromagnetic (EM)
contributions to the cross section for several values of the neutrino
magnetic moment
\cite{Beda:2007hf}.
}
\end{figure}

The two terms $(d\sigma/dT)_{\text{SM}}$ and $(d\sigma/dT)_{\mu}$ exhibit quite different
dependences on the experimentally observable electron kinetic energy $T$,
as illustrated in Fig.~\ref{gemma_f1},
where the values of the two terms
averaged over the typical antineutrino reactor spectrum
are plotted for six values of
the neutrino magnetic moment,
$\mu_{\nu}^{(N)}=N\times10^{-11}\mu_{B}$,
with $N=1,2,3,4,5,6$
\cite{Beda:2007hf}
(see also
\cite{Vogel:1989iv}).
One can see that small values of the neutrino magnetic moment
can be probed by lowering the electron recoil energy threshold.
In fact,
from Eqs.~(\ref{d_sigma_SM}) and (\ref{d_sigma_mu})
one can find that
$(d\sigma/dT)_{\mu}$ exceeds
$(d\sigma/dT)_{\text{SM}}$ for
\begin{equation}
T
\lesssim
\frac{\pi^{2}\alpha^{2}}{G_F^{2}m_{e}^3}
\left(\frac {\mu_{\nu}}{\mu_{B}}\right)^{2}
.
\end{equation}

It was proposed in \cite{Wong:2010pb} that electron binding in atoms (the ``atomic ionization'' effect in
neutrino interactions on Ge target) can significantly increase the
electromagnetic contribution to the differential cross section with
respect to the free electron approximation. However, detailed
considerations of the atomic ionization effect in
(anti)neutrino atomic electron scattering experiments presented in
\cite{Voloshin:2010vm,Kouzakov:2010tx,Kouzakov:2011ig,Kouzakov:2011ka,Kouzakov:2011vx,Kouzakov:2011uq}
show that the effect is by far
too small to have measurable consequences even in the case of the
low energy threshold of 2.8 keV
reached in the GEMMA experiment \cite{Beda:2012-AHEP}.

\subsection{Effective dipole moments}
\label{sec:effective}

In scattering experiments
the neutrino is created at some distance
from the detector as a flavor neutrino,
which is a superposition of massive neutrinos.
Therefore,
the magnetic moment that is measured in these experiment is not that of a massive neutrino,
but it is an effective magnetic moment
which takes into account neutrino mixing and the oscillations
during the propagation between source and detector
\cite{Grimus:1997aa,Beacom:1999wx}:
\begin{equation}
\mu_{\nu}^2(\nu_{\alpha},L,E)
=
\sum_{j}
\left|
\sum_{k}
U_{\alpha k}^{*}
e^{- i m_{k}^{2} L / 2 E }
\left(
\mgm_{jk} - i \elm_{jk}
\right)
\right|^{2}
,
\label{mueff2}
\end{equation}
where we have written explicitly the dependence from
the initial neutrino flavor $\nu_{\alpha}$,
the distance $L$ and the energy $E$.
In this expression of the effective $\mu_{\nu}$
one can see that in general both the magnetic and electric dipole moments
contribute to the elastic scattering.
Note that $\mu_{\nu}(\nu_{\alpha},L,E)$ depends only on the neutrino squared-mass differences:
considering for simplicity only the magnetic moment contribution, we have
\begin{equation}
\mu_{\nu}^2(\nu_{\alpha},L,E)
=
\sum_{j}
\sum_{kk'}
U_{\alpha k}^{*}
U_{\alpha k'}
e^{- i \Delta{m}^{2}_{kk'} L / 2 E }
\mgm_{jk} \mgm_{jk'}
.
\label{mueff3}
\end{equation}

In the case of Majorana neutrinos,
there are no diagonal magnetic and electric dipole moments
and $\mu_{\nu}^2$
receives contributions only from the transition dipole moments.
Furthermore,
if CP is conserved,
there are either only magnetic or electric transition dipole moments
(see Section~\ref{ff:Majorana}).

The general expression for $\mu_{\nu}^{2}(\nu_{\alpha},L,E)$ can
be simplified in some cases
\cite{Beacom:1999wx}.
For instance, for Dirac neutrinos with only diagonal magnetic moments
$\mu_{ij}=\mu_{i}\delta_{ij}$, we have the effective flavor magnetic moment
\begin{equation}
\label{mu_{e}xp_Dir}
\mu_{\nu}^{2}(\nu_{\alpha},L,E)
\rightarrow
(\mu^{\text{D}}_{\alpha})^{2}
=
\sum_{i}
|U_{\alpha i}|^{2} \mu_{i}^{2}
.
\end{equation}
Since in this case there is no dependence on the
distance $L$ and the neutrino energy, the magnetic cross section
is characterized by the initial neutrino flavor rather than by the
composition of mass states in the detector. In this case,
measurements of all flavor magnetic moments and mixing
parameters can allow the extraction of all the fundamental
moments $\mu_{i}$.

\subsection{Experimental limits}
\label{sec:limits}

The constraints on the neutrino magnetic moment in direct
laboratory experiments have been obtained so far from the lack of
any observable distortion of the recoil electron energy spectrum.
Experiments of this type have started about 40 years ago.
The strategy applied during all these years
in reactor experiments is rather simple:
minimize the threshold on the recoil energy for the detection of the scattered electron,
keeping at the same time a reasonable background level. Since the region of interest coincides with the energy region dominated by radioactivity, low 
intrinsic radioactivity detectors have been employed together with active and passive shields. 
In addition, all the experiments were running inside laboratories with significant overburden of concrete, corresponding to several meters of water. This was enough to suppress the soft 
component of cosmic rays.

In all the experiments,
with one exception only, the signal due to the antineutrinos from the reactor
(about $2\times10^{20} \, \text{s}^{-1} \, \text{GWthermal}^{-1}$) is obtained from the difference
between the reactor-on and reactor-off rate.
Clearly, this requires the same background with the reactor on as with the reactor off.

${\bar\nu_{e}}$-$e$
elastic scattering was first observed in the pioneering experiment
\cite{Reines:1976pv} at the Savannah River Laboratory. The set-up was made of
a 15.9 kg plastic scintillator target divided into 16 optically isolated elements, totally enclosed inside a 300 kg 
NaI crystal shielded with lead and cadmium. Finally, the entire set-up was immersed into 2200 liters of liquid scintillator. Both the NaI and the liquid scintillator detectors
were working as veto against cosmics and gamma-rays from the laboratory. In the electron kinetic energy range from 1.5 MeV to 4.5 MeV a reactor-on rate
of 47.5$\pm$1 events/day was measured, to be compared with a reactor-off rate of 40.4$\pm$0.9 events/day. A revised analysis of the Savannah River Laboratory data
\cite{Vogel:1989iv} with an improved reactor neutrino spectrum and a more precise value of sin$^{2}\theta_{W}$ gave hints for a neutrino magnetic moment on
the order of (2-4)$\times 10^{-10} \, \mu_{B}$.

However, lower limits were then obtained by two experiments performed at nuclear reactors in Russia. 
The Krasnoyarsk experiment \cite{Vidyakin:1992nf} had a 103 kg target of liquid organofluoric scintillator 
contained into seven scintillation chambers. The absence of hydrogen in the target was a significant improvement, since the 
${\bar\nu_{e}}$-$p$ charged-current reaction, which has a much larger cross-section than ${\bar\nu_{e}}$-$e$, had been
an important background source in the Savannah River experiment. The active target was then surrounded by a passive shield of steel, copper,
lead and borated polyethylene.
Finally, two layers of plastic scintillators were vetoing the cosmic muons.
With a count rate of 8.3$\pm$0.3 
(reactor-on) and 7.1$\pm$0.4 (reactor-of) in the electron energy range 3.15-5.18 MeV,
the Krasnoyarsk collaboration obtained the limit
\begin{equation}
\mu_{\bar\nu_{e}} \leq 2.4 \times 10^{-10} \, \mu_{B}
\qquad
\text{(90\% C.L.)}
.
\label{lim-Krasnoyarsk}
\end{equation}

A completely different detector was built for the Rovno \cite{Derbin:1993wy}
experiment: 600 silicon detectors, for a total mass of 75 kg, with a passive shield of mercury, copper, cadmium absorber and graphite. Finally, the set-up was enclosed inside a veto made of
plastic scintillators. With a rate of 4963$\pm$12 events/day (reactor-on) and 4921$\pm$16 events/day (reactor-off) in the electron recoil energy range 0.6-2 MeV
it has been possible to obtain a limit on the neutrino magnetic moment of
\begin{equation}
\mu_{\bar\nu_{e}} \leq 1.9 \times 10^{-10} \, \mu_{B}
\qquad
\text{(90\% C.L.)}
.
\label{lim-Rovno}
\end{equation}

Finally, more stringent limits have been obtained in the two most recent experiments at rectors.
TEXONO \cite{Wong:2006nx} has been performed at the Kuo-Sheng nuclear power station. The detector, a 1.06 kg high purity germanium, 
was completely surrounded by NaI (Tl) and CsI(Tl) crystals working as anti-Compton. The whole set-up was contained inside a shield made of copper,
boron loaded polyethylene, stainless steel, lead and plastic scintillators. A background of about 1 event/keV$\cdot$kg$\cdot$day could be achieved above the threshold of 12 keV,
giving the limit
\begin{equation}
\mu_{\bar\nu_{e}} \leq 7.4 \times 10^{-11} \, \mu_{B}
\qquad
\text{(90\% C.L.)}
.
\label{lim-TEXONO}
\end{equation}

At the moment the world best limit is coming from the GEMMA experiment at the Kalinin nuclear power plant.
A 1.5 kg high purity germanium detector is placed inside a cup-shaped NaI crystal and surrounded by copper, lead and plastic scintillators.
With
an energy threshold as low as 2.8 keV,
the GEMMA collaboration obtained
\cite{Beda:2012-AHEP}
\begin{equation}
\mu_{\bar\nu_{e}} \leq 2.9 \times 10^{-11} \, \mu_{B}
\qquad
\text{(90\% C.L.)}
.
\label{lim-GEMMA}
\end{equation}

The experiment which followed a strategy different from the study of the reactor-on and reactor-off rate was MUNU.
As a matter of fact, the detector \cite{Amsler:1997pn} was able to provide not only the energy but also the topology of events. As a consequence, the initial direction of an 
electron track could be measured and the electron scattering angle reconstructed. This allowed to look for the reactor signal by 
comparing forward electrons, having as reference the reactor to detector axis, with the backward ones.
In this way, the background is measured on-line, which eliminates problems 
from detector instabilities, as well as from a possible time dependence of the background itself. The central component of the 
detector consisted of an acrylic vessel time projected chamber (a cylinder 90 cm in diameter and 162 cm long) filled with CF$_{4}$ at 3 bar pressure and
immersed in a steel tank (2 m diameter and 3.8 m long) filled with 10 m$^{3}$ liquid scintillator viewed by 48 photomultipliers. 
The total target mass of CF$_{4}$ was 11.4 kg. Finally, the set-up was surrounded by boron loaded polyethylene and lead. With a total rate of 6.8$\pm$0.3 events/day in
the forward direction and a background of 5.8$\pm$0.17 events/day
the following upper bounds have been obtained 
\cite{Daraktchieva:2005kn}:
\begin{equation}
\mu_{\bar\nu_{e}} \leq 9 \times 10^{-11} \, \mu_{B}
\qquad
\text{(90\% C.L.)}
.
\label{lim-MUNU}
\end{equation}

Several experiments at accelerators have searched for an effect due to the magnetic moment of $\nu_{\mu}$
in
$\nu_{\mu}$--$e$
and
$\bar\nu_{\mu}$--$e$
elastic scattering
(see \cite{PDG-2012}).
The current best limit has been obtained in the LSND experiment
\cite{Auerbach:2001wg}:
\begin{equation}
\mu_{\nu_{\mu}} \leq 6.8 \times 10^{-10} \, \mu_{B}
\qquad
\text{(90\% C.L.)}
.
\label{lim-LSND}
\end{equation}

The DONUT collaboration have investigated
$\nu_{\tau}$--$e$
and
$\bar\nu_{\tau}$--$e$
elastic scattering,
finding the limit
\cite{Schwienhorst:2001sj}
\begin{equation}
\mu_{\nu_{\tau}} \leq 3.9 \times 10^{-7} \, \mu_{B}
\qquad
\text{(90\% C.L.)}
.
\label{lim-DONUT}
\end{equation}

Solar neutrino experiments as Super-Kamiokande and Borexino can also search for a neutrino magnetic moment signal by
studying the shape of the electron spectrum. Since the neutrino magnetic
moment depends both on the mixing and on the 
propagation properties of the neutrino then oscillations are here relevant. 

The analysis of the recoil electron spectrum generated by solar neutrinos in the Super-Kamiokande experiment
experiment gave
\cite{Liu:2004ny}:
\begin{equation}
\mu_{\nu} \leq 1.1 \times 10^{-10} \, \mu_{B}
\qquad
\text{(90\% C.L.)}
,
\label{lim-SK}
\end{equation}
where $\mu_{\nu}$ is not the same as $\mu_{\bar\nu_{e}}$
since it is given by a different combination of the magnetic moment of the neutrino mass eigenstates
(see Section~\ref{sec:effective}).

The limit
\begin{equation}
\mu_{\nu} \leq 5.4 \times 10^{-11}\mu_{B}
\qquad
\text{(90\% C.L.)}
,
\label{lim-Borexino-el}
\end{equation}
has been recently obtained in the
Borexino solar neutrino scattering experiment
\cite{Arpesella:2008mt}.
An upper limit on the neutrino magnetic moment
$\mu_{\nu} \leq 8.4 \times 10^{-11}\mu_{B}$
has been found in an independent analysis
of the first release of the
Borexino experiment data performed in \cite{Montanino:2008hu}.
It was also shown that with reasonable assumptions on the
oscillation probability this limit translates into the conservative upper
limits on the magnetic moments of $\nu_\mu$ and $\nu_\tau$:
\begin{equation}
\mu_{\nu_\mu} \leq 1.5 \times 10^{-10} \, \mu_{B}
,
\qquad
\mu_{\nu_\tau} \leq 1.9 \times 10^{-10} \, \mu_{B}
\qquad
\text{(90\% C.L.)}
.
\label{lim-Borexino-mu-tau}
\end{equation}
The limit on $\mu_{\nu_\tau}$ is three order of magnitude stronger than
the direct limit in Eq.~(\ref{lim-DONUT}).

The global fit \cite{Grimus:2002vb,Tortola:2004vh} of the magnetic moment data from the reactor and
solar neutrino experiments for the Majorana neutrinos produces
limits on the neutrino transition moments
\begin{equation}
\mu_{23},\mu_{31},\mu_{12} < 1.8 \times 10^{-10} \mu_{B}
\qquad
\text{(90\% C.L.)}
.
\label{lim-transition}
\end{equation}

Finally, an interesting new possibility for providing more stringent
constraints on the neutrino magnetic moment from $\bar \nu_{e}$-$e$
scattering experiments was discussed in \cite{Bernabeu:2004ay} on
the basis of an observation \cite{Segura:1993tu} that
``dynamical zeros" appear in the Standard Model contribution to
the scattering cross section.

\subsection{Theoretical considerations}
\label{sec:Theoretical}

As it was already mentioned before, there is a gap of many orders
of magnitude between the present experimental limits
$\sim10^{-11}\mu_{B}$
on neutrino magnetic moments
(discussed in Section~\ref{sec:limits})
and the prediction (\ref{mu_3_10_19}) of the
minimal extension of the Standard Model with right-handed neutrinos.
At the same time, the
experimental sensitivity
of reactor ${\bar\nu_{e}}$-$e$
elastic scattering experiments
have improved by only one order of
magnitude during a period of about twenty years
(see \cite{Vogel:1989iv}, where a sensitivity of $\sim10^{-10}\mu_{B}$ is discussed).
However,
the experimental studies of neutrino magnetic moments
are stimulated by the hope that
new physics beyond the minimally extended Standard Model with right-handed neutrinos might
give much stronger contributions. One of the
examples in which it is possible to avoid the neutrino magnetic
moment being proportional to a (small) neutrino mass, that would
in principle make a neutrino magnetic moment accessible for
experimental observations, is realized in the left-right symmetric
models considered at the end of Section~\ref{sec:theoretical}.

Other interesting possibilities of obtaining neutrino magnetic
moments lager than the prediction (\ref{mu_3_10_19}) of the
minimal extension the Standard Model with right-handed neutrinos have been considered
recently. In this concern, we note that it was proposed in
\cite{Mohapatra:2004ce} to probe a class of large extra dimensions
models with future reactors searches for neutrino magnetic
moments. The results obtained within the Minimal Supersymmetric
Standard Model with $R$-parity violating interactions
\cite{Gozdz:2006iz,Gozdz:2006jv} show that the Majorana transition
magnetic moment might be significantly above the scale of
(\ref{mu_3_10_19}).

Considering the problem of
large neutrino magnetic moments, one can write down
a generic
relation between the size of a neutrino magnetic moment $\mu_{\nu}$ and the
corresponding neutrino mass
$m_{\nu}$
\cite{Voloshin:1987qy,Barr:1990um,Pal:1991pm, Bell:2005kz,Bell:2006wi,Bell:2007nu}.
Suppose that a large neutrino magnetic moment is
generated by physics beyond a minimal extension of the Standard
Model at an energy scale characterized by $\Lambda$. For a
generic diagram corresponding to this contribution to $\mu_{\nu}$,
one can again use the Feynman graph in Fig.~\ref{A004}; the shaded
circle in this case denotes effects of new physics beyond the
Standard Model. The contribution of this diagram to the magnetic
moment is
\begin{equation}\label{mu_Lambda}
\mu_{\nu} \sim \frac{eG}{\Lambda},
\end{equation}
where $e$ is the electric charge and $G$ is a combination of coupling
constants and loop factors. The same diagram of Fig.~\ref{A004}
but
without the photon line gives a new physics contribution to the
neutrino mass
\begin{equation}\label{m_Lambda}
\delta m_{\nu} \sim G\Lambda.
\end{equation}
Combining the estimates (\ref{mu_Lambda}) and (\ref{m_Lambda}), one
can get the relation
\begin{equation}\label{mu_Lambda1}
\delta m_{\nu} \sim \frac{\Lambda^{2}}{2m_{e}}\frac{\mu_{\nu}}{\mu_{B}}=
\frac{\mu_{\nu}}{10^{-18}\mu_{B}}\left(\frac{\Lambda}{1 \, \text{TeV}}\right)^{2}\
\text{eV}
\end{equation}
between the one-loop contribution to the neutrino mass and the
neutrino magnetic moment.

It follows that, generally, in theoretical models that predict
large values for the neutrino magnetic moment, simultaneously
large contributions to the neutrino mass arise. Therefore, a
particular fine tuning is needed to get a large value for the
neutrino magnetic moment while keeping the neutrino mass within
experimental bounds. One of the possibilities
\cite{Voloshin:1987qy} is based on the idea of suppressing the
ratio $m_{\nu}/\mu_{\nu}$ with a symmetry: if a $SU(2)_{\nu}$
symmetry is an exact symmetry of the Lagrangian of a model,
because of different symmetry properties of the mass and magnetic
moment even a massless neutrino can have a nonzero magnetic
moment. If, as it happens in a realistic model, the $SU(2)_{\nu}$
symmetry is broken and if this breaking is small, the ratio
$m_{\nu}/\mu_{\nu}$ is also small, giving a natural way to obtain
a magnetic moment of the order of $\sim 10^{-11}\mu_{B}$ without
contradictions with the neutrino mass experimental constraints.
Several possibilities based on the general idea of
\cite{Voloshin:1987qy}
were considered in
\cite{Leurer:1989hx,Babu:1990wv,Georgi:1990se,Ecker:1989ph,Chang:1991ri,Barbieri:1988fh}.

Another idea of neutrino mass suppression without suppression of
the neutrino magnetic moment was discussed in \cite{Barr:1990um}
within the Zee model \cite{Zee:1980ai}, which is based on the
Standard Model gauge group $SU(2)_{L}\times U(1)_{Y}$ and
contains at least three Higgs doublets and a charged field which
is a singlet of $SU(2)_{L}$. For this kind of models there is a
suppression of the neutrino mass diagram, while the magnetic
moment diagram is not suppressed.

It is possible to show with more general and rigorous
considerations \cite{Bell:2005kz,Bell:2006wi,Bell:2007nu} that the
${\Lambda}^{2}$ dependence in Eq.~(\ref{mu_Lambda1}) arises from the
quadratic divergence in the renormalization of the dimension-four
neutrino mass operator. A general and model-independent upper
bound on the Dirac neutrino magnetic moment, which can be
generated by an effective theory beyond the Standard Model, has
been derived \cite{Bell:2005kz,Bell:2006wi,Bell:2007nu} from the
demand of absence of fine-tuning of effective operator
coefficients and from the current experimental information on
neutrino masses. A model with Dirac fermions, scalars and gauge
bosons that is valid below the scale $\Lambda$ and respects the
Standard Model $SU(2)_L \times U(1)_Y$ symmetry was considered.
Integrating out the physics above the scale $\Lambda$, the
following effective Lagrangian that involves right-handed
neutrinos $\nu_{R}$, lepton isodoublets and the Higgs doublet can
be obtained:
\begin{equation}\label{L_operator}
\mathcal{L}_{eff}=\sum_{n,j} \frac{\mathcal{C}^{n}_{j}(\mu)}{\Lambda
^{n-4}}\mathcal{O}_{j}^{(n)}(\mu)
+ \text{H.c.}
,
\end{equation}
where $\mu$ is the renormalization scale, $n\geq 4$ denotes the
operator dimension and $j$ runs over independent operators of a
given dimension. For $n=4$, a neutrino mass arises from the operator
$\mathcal{O}^{(4)}_{1}={\bar L}{\tilde \Phi}\nu_{R}$, where
${\tilde \Phi}=i\sigma_2 \Phi^{*}$. In addition, if the scale
$\Lambda$ is not extremely large with respect to the electroweak
scale, an important contribution to the neutrino mass can
arise also from higher dimension operators. At this point it
is important to note that the combination of the $n=6$ operators
appearing in the Lagrangian (\ref{L_operator}) contains the
magnetic moment operator ${\bar \nu}\sigma_{\mu \nu} \nu F^{\mu
\nu}$ and also generates a contribution $\delta m_{\nu}$ to the
neutrino mass
\cite{Bell:2005kz,Bell:2006wi,Bell:2007nu}. Solving the
renormalization group equation from the scale $\Lambda$ to the
electroweak scale, one finds that the contributions to the
neutrino magnetic moment and to the neutrino mass are connected to
each other by
\begin{equation}\label{mu_D_Bell}
|\mu_{\nu}^{\text{D}}|=\frac{16{\sqrt 2}G_Fm_{e} \delta m_{\nu}\sin^{4}\theta
_{W}}{9\alpha^{2}|f|\ln\left(\Lambda / {{\it v}}\right)}\mu_{B},
\end{equation}
where $\alpha$ is the fine structure constant, ${\it v}$ is the
vacuum expectation value of the Higgs doublet,
\begin{equation}\label{f_r}
f=1-r-\frac{2}{3}\tan^{2} \theta_{W} -\frac{1}{3}(1+r)\tan^4 \theta_{W},
\end{equation}
and $r$ is a ratio of effective operator coefficients defined at the scale
$\Lambda$ which is of order unity without fine-tuning. If the neutrino magnetic
moment is generated by new physics at a scale $\Lambda \sim 1 \ \text{TeV}$ and
the corresponding contribution to the neutrino mass is $\delta m_{\nu} \lesssim
1\ \text{eV}$, then the bound $\mu_{\nu}\lesssim 10^{-14} \mu_{B}$ can be
obtained. This bound is some orders of magnitude stronger than the
constraints from reactor and solar neutrino scattering experiments discussed
before.

The model-independent limit on a Majorana neutrino transition
magnetic moment $\mu^{\text{M}}_{\nu}$ was also discussed in
\cite{Bell:2005kz,Bell:2006wi,Bell:2007nu}. However, the limit in
the Majorana case is much weaker than that in the Dirac case,
because for a Majorana neutrino the magnetic moment contribution
to the mass is Yukawa suppressed. The limit on $\mu^{\text{M}}_{\nu}$ is
also weaker than the present experimental limits if
$\mu^{\text{M}}_{\nu}$ is generated by new physics at the scale $\Lambda \sim
1 \ \text{TeV}$. An important conclusion of
\cite{Bell:2005kz,Bell:2006wi,Bell:2007nu}, based on
model-independent considerations of the contributions to
$\mu_{\nu}$, is that if a neutrino magnetic moment of order $\mu
_{\nu} \geq 10^{-15} \mu_{B}$ were observed in an experiment, it
would give a proof that neutrinos are Majorana rather than Dirac
particles.

\section{Neutrino charge radius}
\label{sec5}

Even if the electric charge of a neutrino is vanishing, the electric form factor
$f_Q(q^{2})$ can still contain nontrivial information about neutrino electromagnetic
properties.
Considering $f_{Q}(0)=0$,
in the static limit ($q^{2}\to0$),
the electric form factor is given by
\begin{equation}\label{nu_cha_rad}
f_{Q}(q^{2})
=
q^{2}
\left.
\frac{df_{Q}(q^{2})}{dq^{2}}
\right|_{q^{2}=0}
+
\ldots
.
\end{equation}
The leading contribution can be expressed in terms of a
neutrino charge radius considering a static spherically symmetric charge
distribution of density $\rho(r)$ (with $r=|{\vec{x}}|$)
in the so-called
``Breit frame'', where $q_0=0$.
In this approximation, we have
\begin{equation}
f_{Q}(q^{2})
=
\int
\rho(r)
e^{i\vet{q}\cdot\vet{x}}
d^3x
=
4\pi
\int
\rho(r)
\frac{\sin(qr)}{qr}
r^{2} dr
,
\end{equation}
where $q=|{\vet{q}}|$.
Since
$
\left.
df_{Q}/dq^{2}
\right|_{q^{2}=0}
=
-
\langle{r}^{2}\rangle / 6
$,
with
$
\langle{r}^{2}\rangle
=
\int
r^2
\rho(r)
e^{i{\vet{q}}{\vet{x}}}
d^3x
$,
the neutrino charge radius
is defined by
\begin{equation}
\label{nu_cha_rad_1}
{\langle{r}_{\nu}^{2}\rangle}
=
-
6
\left.
\frac{df_{Q}(q^{2})}{dq^{2}}
\right|_{q^{2}=0}
.
\end{equation}
Note that
${\langle{r}_{\nu}^{2}\rangle}$ can be negative,
because the charge density $\rho(r)$ is not a positively defined function of $r$.

In one of the first studies \cite{Bardeen:1972vi}, it was
claimed that in the Standard Model and in the unitary gauge the
neutrino charge radius is ultraviolet-divergent and so it is not a
physical quantity.
A direct one-loop calculation
\cite{Dvornikov:2003js, Dvornikov:2004sj} of proper vertices
%(Fig.~\ref{fig1f})
and $\gamma - Z$ self-energy
%(Fig.~\ref{fig4_0})
(Figs.~(15) and (16) of \cite{Giunti:2008ve})
contributions to the neutrino charge radius
performed in a general $R_\xi $ gauge for a massive Dirac
neutrino gave also a divergent result. However, it was shown
\cite{Lee:1973fw}, using the unitary gauge, that by including in
addition to the usual terms also contributions from diagrams of
the neutrino-lepton neutral current scattering ($Z$ boson
diagrams), it is possible to obtain for the neutrino charge radius
a gauge-dependent but finite quantity.
Later on, it was also shown
\cite{Lee:1977tib} that in order to define the neutrino charge
radius as a physical quantity one has also to consider box
diagrams
(see Fig.~(18) of \cite{Giunti:2008ve}),
which contribute to the scattering process
$\nu+\ell\to\nu+\ell$,
and that in combination with contributions from the
proper diagrams it is possible to obtain a finite and
gauge-independent value for the neutrino charge radius. In this
way, the neutrino electroweak radius was defined
\cite{Lucio:1983mg,Lucio:1984jn} and an additional set of diagrams
that give contribution to its value was discussed in
\cite{Degrassi:1989ip}. Finally, in a series of papers
\cite{Bernabeu:2000hf,Bernabeu:2002nw,Bernabeu:2002pd} the
neutrino electroweak radius as a physical observable has been
introduced. In the corresponding calculations, performed in the
one-loop approximation including additional terms from the
$\gamma-Z$ boson mixing and the box diagrams involving $W$ and $Z$
bosons, the following gauge-invariant result for the neutrino
charge radius have been obtained:
\begin{equation}
{\langle
r_{\nu_{\alpha}}^{2}\rangle}
=
\frac{G_F}{4\sqrt{2}\pi^{2}}
\left[
3-2\log\left(\frac{m_{\alpha}^{2}}{m^{2}_{W}}\right)
\right]
,
\end{equation}
where $m_{W}$ and $m_{\alpha}$ are the $W$ boson and lepton masses
($\alpha=e,\mu,\tau$).
This result, however, revived the discussion
\cite{Fujikawa:2003tz,Fujikawa:2003ww,Papavassiliou:2003rx,Bernabeu:2004jr}
on the definition of the neutrino charge radius.
Numerically, for the electron neutrino electroweak radius it
yields
\cite{Bernabeu:2000hf,Bernabeu:2002nw,Bernabeu:2002pd}
\begin{equation}
{\langle{r}_{\nu_{e}}^{2}\rangle}=4 \times 10^{-33} \, \text{cm}^{2},
\label{r00}
\end{equation}
which is very close to the numerical estimations obtained much
earlier in \cite{Lucio:1983mg,Lucio:1984jn}.

Note that the neutrino charge radius can be considered as an
effective scale of the particle's ``size'', which should influence
physical processes such as, for instance, neutrino scattering off
electron.
To incorporate the neutrino charge
radius contribution in the cross section (\ref{d_sigma_SM}), the following
substitution \cite{Grau:1985cn,Vogel:1989iv,Hagiwara:1994pw} can
be used:
\begin {equation}\label{g_V_ch_rad}
g_V\rightarrow \frac{1}{2}+2\sin^{2} \theta_{W} + \frac{2}{3}m^{2}_{W}
{\langle{r}_{\nu_{e}}^{2}\rangle}\sin^{2} \theta_{W}.
\end{equation}
Using this method,
the TEXONO collaboration obtained
\cite{Deniz:2009mu}
\begin{equation}
-2.1\times 10^{-32} \, \text{cm}^{2}
<
\langle{r}_{\bar\nu_{e}}^{2}\rangle
<
3.3 \times 10^{-32} \, \text{cm}^{2}
\qquad
\text{(90\% C.L.)}
.
\label{r01}
\end{equation}

Other available bounds on the electron neutrino charge radius are:
from primordial nucleosynthesis
\cite{Grifols:1986ed}
\begin{equation}
\langle{r}_{\nu_{e}}^{2}\rangle
\lesssim
7 \times 10^{-33} \, \text{cm}^{2}
,
\label{r02}
\end{equation}
from SN 1987A \cite{Grifols:1989vi}
\begin {equation}
\langle{r}_{\bar\nu_{e}}^{2}\rangle
\lesssim
2 \times 10^{-33} \, \text{cm}^{2}
,
\label{r03}
\end{equation}
from neutrino neutral-current reactions \cite{Allen:1990xn}
\begin {equation}
-2.74\times 10^{-32} \, \text{cm}^{2}
<
\langle{r}_{\nu_{e}}^{2}\rangle
<
4.88 \times 10^{-32} \, \text{cm}^{2}
\qquad
\text{(90\% C.L.)}
,
\label{r04}
\end{equation}
from solar experiments (Kamiokande II and Homestake) \cite{Mourao:1992ip}
\begin{equation}
\langle{r}_{\nu_{e}}^{2}\rangle < 2.3 \times 10^{-32} \, \text{cm}^{2}
\qquad
\text{(95\% C.L.)}
,
\label{r05}
\end{equation}
from an evaluation of the weak mixing angle $\sin^{2} \theta_{W}$ by a
combined fit of all electron neutrino elastic scattering data
\cite{Barranco:2007ea}
\begin {equation}
-0.13\times 10^{-32} \, \text{cm}^{2}
<
\langle{r}_{\nu_{e}}^{2}\rangle
<
3.32 \times 10^{-32} \, \text{cm}^{2}
\qquad
\text{(90\% C.L.)}
.
\label{r06}
\end{equation}

Comparing the theoretical value in Eq.~(\ref{r00})
with the experimental limits in Eqs.~(\ref{r01})--(\ref{r06}),
one can see that they differ at most by one
order of magnitude.
Therefore, one may expect that the experimental
accuracy will soon reach the value needed to probe the theoretical predictions for the neutrino
effective charge radius.

The effects of new physics beyond the Standard
Model can also contribute to the neutrino charge radius.
let us only mention that
the anomalous $WW\gamma$ vertex contribution to
the neutrino effective charge radius has been studied in \cite{NovalesSanchez:2008tn},
and shown to correspond to a contribution
$\lesssim 10^{-34} \, \text{cm}^{2}$
to
$|\langle{r}_{\nu_{e}}^{2}\rangle|$.
Note that this is only one order of magnitude lower
than the expected value of the charge radius in the Standard Model.

A detailed discussion on the possibilities to constrain the $\nu_{\tau}$
and $\nu_{\mu}$ charge radii from astrophysical and cosmological
observations and from terrestrial experiments can be found in
\cite{Hirsch:2002uv}.
\section{Radiative decay and plasmon decay}
\label{sec6}

If the masses of neutrinos are non-degenerate, the radiative decay of
a heavier neutrino $\nu_{i}$ into a lighter neutrino $\nu_{f}$ (with $m_{i}>m_{f}$ ) with
emission of a photon,
\begin{equation}
\nu_{i} \to\nu_{f} + \gamma
,
\label{eq:radiativedecay}
\end{equation}
may proceed in vacuum
\cite{Marciano:1977wx,Lee:1977tib,Petcov:1976ff,Goldman:1977jx,Bilenky:1987ty,Zatsepin:1978iy,Pal:1981rm}.
Early discussions of the possible role of neutrino
radiative decay in different astrophysical and cosmological
settings can be found in
\cite{Dicus:1977nn,Sato:1977ye,Stecker:1980bu,Kimble:1980vz,Melott:1981iw,DeRujula:1980qd}.

For the case of a Dirac neutrino, the decay rate
in the minimal extension of the
Standard Model with right-handed neutrinos
is
\cite{Marciano:1977wx,Lee:1977tib,Petcov:1976ff,Goldman:1977jx,Bilenky:1987ty,Zatsepin:1978iy,Pal:1981rm}
\begin{equation}
\Gamma_{\nu^D_{i}\to\nu^D_{j}+\gamma}
=
\frac{\alpha G_F^{2}}{128 \pi^4}
\left(\frac{m^{2}_i-m^{2}_j}{m_j}\right)^3
\left(m^{2}_i+m^{2}_j\right)
\left|
\sum_{l=e,\mu,\tau}
f(a_l) U_{lj} U^{*}_{li} \right|^{2}
,
\label{Gamma_Dir_nu_i+nu_j+gamma}
\end{equation}
where $f(a_l)$ is given by Eq.~(\ref{f_a_l}). Recalling the results
for the Dirac neutrino magnetic and electric
transition moments $\mu_{ij}$ and $\epsilon_{ij}$, given in
Eq.~(\ref{m_e_mom_i_j}), one can rewrite
Eq.~(\ref{Gamma_Dir_nu_i+nu_j+gamma}) as
(see \cite{Raffelt:1996wa,Raffelt:1999gv})
\begin{equation}
\label{Gamma}
\Gamma_{\nu_{i}\to\nu_{j}+\gamma}
=
\frac{|\mu_{ij}|^{2}+|\epsilon_{ij}^{2}|} {8 \pi}\left(\frac{m^{2}_i-m^{2}_j}{m_j}\right)^3
.
\end{equation}
For degenerate neutrino masses ($m_i = m_j$), the process is
kinematically forbidden in vacuum.

Note that there are models
(see for instance \cite{Petcov:1982en})
in which the neutrino radiative decay rate
(as well as the magnetic moment discussed above)
of a non-standard
Dirac neutrinos are much larger than those predicted in the minimally
extended Standard Model.

For Majorana neutrinos,
if CP is violated the decay rate is given by Eq.~(\ref{Gamma}).
If CP is conserved we have two cases:
If the Majorana neutrinos $\nu_i$ and $\nu_j $ have the same CP eigenvalues,
Eqs.~(\ref{A050}) and (\ref{edm:Majorana}) imply that the decay process is induced purely by the neutrino
electric transition dipole moment, because $\mu_{ij}=0$;
on the other hand,
if the two Majorana neutrinos have opposite CP eigenvalues,
from
Eqs.~(\ref{A051}) and (\ref{mdm:Majorana})
one can see that the transition is purely of magnetic dipole type ($\epsilon_{ij}=0$).

For numerical estimations it is convenient to express Eq.~(\ref{Gamma})
in the following form:
\begin{equation}\label{Gamma_num}
\Gamma_{\nu_{i}\to\nu_{j}+\gamma}
=
5.3
\left(\frac{\mu_{\text{eff}}}{\mu_{B}}\right)^{2}
\left(\frac{m^{2}_i-m^{2}_j}{m_j^{2}}\right)^3
\left(\frac{m_i}{1\ \text{eV}}\right)^3 s^{-1},
\end{equation}
with the effective neutrino magnetic moment
$\mu_{\text{eff}}=\sqrt{|\mu_{ij}|^{2}+|\epsilon_{ij}^{2}|}$.

\begin{figure}
\begin{center}
\includegraphics*[width=0.5\linewidth]{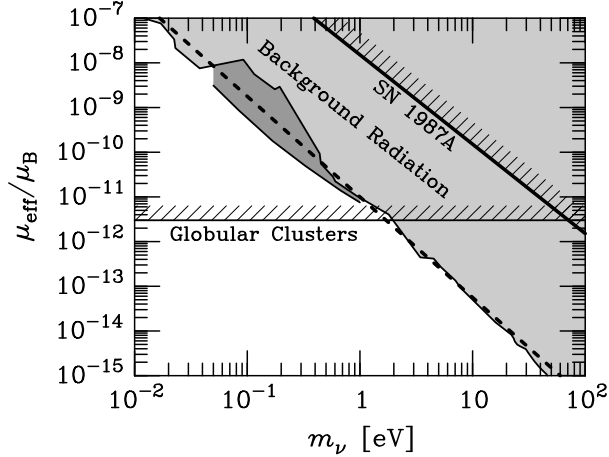}
\end{center}
\caption{\label{Raffelt-PhysRept-320-319-1999-f2}
Astrophysical limits on neutrino transition moments \cite{Raffelt:1996wa,Raffelt:1999gv}.
}
\end{figure}

The neutrino radiative decay can be constrained by the absence of
decay photons in reactor $\bar\nu_{e}$ and solar $\nu_{e}$ fluxes.
The limits on $\mu_{\text{eff}}$ that have been obtained from these
considerations are much weaker than those obtained from neutrino
scattering terrestrial experiments. Stronger constraints on
$\mu_{\text{eff}}$ (though still weaker than the terrestrial ones) have been
obtained from the neutrino decay limit set by SN 1987A and
from the limits on the distortion of the Cosmic Microwave Background Radiation (CMBR).
These limits can be expressed as (see
\cite{Raffelt:1996wa,Raffelt:1999gv} and references therein)
\begin{equation}
\frac{\mu_{\rm eff}}{\mu_{\rm B}}
<
\left\{\begin{array}{ll}
0.9{\times}10^{-1} (\text{eV}/m_{\nu})^{2} & \text{Reactor ($\bar\nu_{e}$)},\\
0.5{\times}10^{-5} (\text{eV}/m_{\nu})^{2} & \text{Sun ($\nu_{e}$)}, \\
1.5{\times}10^{-8} (\text{eV}/m_{\nu})^{2} & \text{SN 1987A (all flavors)}, \\
1.0{\times}10^{-11}(\text{eV}/m_{\nu})^{9/4} & \text{CMBR (all flavors)}.
\end{array}\right.
\end{equation}
Detailed discussions (and corresponding references) on the astrophysical
constraints on the neutrino magnetic and electric transition moments,
summarized in Fig.~\ref{Raffelt-PhysRept-320-319-1999-f2},
can be found in \cite{Raffelt:1996wa,Raffelt:1999gv}.

For completeness, we would like to mention that other processes
characterized by the same signature of Eq.~(\ref{eq:radiativedecay})
have been considered (for a review of the literature
see \cite{Ioannisian:1996pn,Lobanov:2002ur,Studenikin:2004bu,Studenikin:2007zz,Studenikin:2006jr}):

\begin{enumerate}

\renewcommand{\labelenumi}{\theenumi)}
\renewcommand{\theenumi}{\roman{enumi}}

\item
The photon radiation by a massless neutrino
$(\nu_{i} \to\nu_{j} + \gamma,\; i=j)$ due to the vacuum
polarization loop diagram in the presence of an external magnetic
field \cite{GalNik:1972at}.

\item
The photon radiation by a massive neutrino with nonvanishing
magnetic moment in constant magnetic and electromagnetic wave
fields
\cite{Borisov:1988wy,Borisov:1989yw,Skobelev:1976at,Skobelev:1991pv}.

\item
The Cherenkov radiation due to the nonvanishing neutrino
magnetic moment in an homogeneous and infinitely extended medium,
which is only possible if the speed of the neutrino is larger than
the speed of light in the medium
\cite{Radomski:1975re,Grimus:1993uw}.

\item
The transition radiation due to a nonvanishing neutrino
magnetic moment which would be produced when the neutrino crosses
the interface of two media with different refractive indices
\cite{Sakuda:1993aq,Sakuda:1994zq,Grimus:1994ug}.

\item
The Cherenkov radiation of a massless neutrino due to its
induced charge in a medium
\cite{Oraevsky:1986dt,D'Olivo:1995gy}\footnote {Note that the
neutrino electromagnetic properties are in general affected by the
external environment. In particular, a neutrino can acquire an
electric charge in magnetized matter
\cite{Oraevsky:1986dt,D'Olivo:1995gy} and the neutrino magnetic
moment depends on the strength of external electromagnetic fields
\cite{Borisov:1985ha,Borisov:1989yw,Masood:1999qv}.
A recent study of the neutrino electromagnetic vertex in
magnetized matter can be found in \cite{Nieves:2003kw}.
See also
\cite{Studenikin:2004bu,Studenikin:2007zz} for a review of
neutrino interactions in external electromagnetic fields.}.

\item
The Cherenkov radiation of massive and massless neutrinos in
a magnetized medium \cite{Mohanty:1995bx, Ioannisian:1996pn}.

\item
The neutrino radiative decay $(\nu_{i} \to\nu_{j} +
\gamma, \; i\not=j)$ in external fields and media (see
\cite{Giunti:1990pp,Gvozdev:1992bh,Skobelev:1995pf,Zhukovsky:1996bi,Kachelriess:1996up,Ternov:2003yi}
and references therein).

\item
The spin light of neutrino in matter ($SL\nu$) that is a
mechanism of electromagnetic radiation due to the precession or
transition of magnetic or electric (transition) moments
of massive neutrinos when they propagate in background matter
\cite{Lobanov:2002ur,Lobanov:2004um,Grigoriev:2004bm,Studenikin:2004dx,Grigorev:2005sw,Lobanov:2005zn}.

\end{enumerate}

A very interesting process, for the purpose of constraining
neutrino electromagnetic properties, is the photon (plasmon) decay
into a neutrino-antineutrino pair:
\begin{equation}
\gamma^{*} \to \nu + \bar\nu
.
\label{plasmon}
\end{equation}
This process becomes kinematically allowed
in media, because a photon with the dispersion relation
$\omega_{\gamma}^{2} + {\vec k}_{\gamma}^{2} >0$ roughly behaves as
a particle with an effective mass.

Plasmon decay
generated by the neutrino coupling
to photons due to a magnetic moment $\mu_{\nu}$
(and/or to a neutrino electric millicharge $q_{\nu}$)
was first considered
in \cite{Bernstein:1963qh} as a possible source of energy loss of the Sun.
The requirement that the energy loss does not
exceed the solar luminosity,
gave
\cite{Raffelt:1996wa,Raffelt:1999gv}
\begin{equation}
\mu_{\nu} \lesssim 4 \times 10^{-10} \, \mu_{B}
,
\label{p01}
\end{equation}
and
$q_{\nu}\lesssim 6 \times 10^{-14} e$.

The tightest astrophysical bound on a neutrino magnetic moment is provided by
the observed properties of globular cluster stars.
The plasmon decay (\ref{plasmon}) inside the star liberates the energy
$\omega_{\gamma}$ in the form of neutrinos that freely escape the stellar
environment. This nonstandard energy loss cools a red giant star so fast that
it can delay helium ignition.
From the lack of observational evidence of this effect,
the following limit has been found \cite{Raffelt:1990pj}:
\begin{equation}
\mu_{\nu}\leq 3 \times 10^{-12} \, \mu_{B}
,
\label{p02}
\end{equation}
and
$q_{\nu}\lesssim 2 \times 10^{-14} e$.
This is the most stringent astrophysical constraint on a neutrino
magnetic moment, applicable to both Dirac and Majorana neutrinos. The
same limit applies for the neutrino magnetic transition moments as well
as for the electric (transition) moments.

Recently,
it has been shown that
the additional cooling due to neutrino magnetic moments
generates qualitative changes to the structure and evolution of stars
with masses between 7 and 18 solar masses,
rather than simply changing the time scales of their burning
\cite{Heger:2008er}.
The resulting sensitivity to the neutrino magnetic moment
has been estimated to be at the level of $(2-4) \times 10^{-11} \, \mu_B$.
\section{Spin-flavor precession}
\label{sec7}

If neutrinos have magnetic moments, the spin can precess in a transverse magnetic field
\cite{Cisneros:1971nq,Voloshin:1986ty,Okun:1986na}.

Let us first consider the spin precession of a Dirac neutrino generated by its diagonal magnetic moment
$\mgm$.
The spatial evolution of the left-handed
and right-handed
helicity amplitudes
$\varphi_{L}(x)$
and
$\varphi_{R}(x)$
in a transverse magnetic field
$B_{\perp}(x)$ is given by
\begin{equation}
i \frac{ \text{d} }{ \text{d}x }
\begin{pmatrix}
\varphi_{L}(x)
\\
\varphi_{R}(x)
\end{pmatrix}
=
\begin{pmatrix}
0 & \mu B_{\perp}(x)
\\
\mu B_{\perp}(x) & 0
\end{pmatrix}
\begin{pmatrix}
\varphi_{L}(x)
\\
\varphi_{R}(x)
\end{pmatrix}
. \label{102}
\end{equation}
The differential equation (\ref{102}) can be solved through the
transformation
\begin{equation}
\begin{pmatrix}
\varphi_{L}(x)
\\
\varphi_{R}(x)
\end{pmatrix}
= \frac{1}{\sqrt{2}}
\begin{pmatrix}
1 & 1
\\
-1 & 1
\end{pmatrix}
\begin{pmatrix}
\varphi_{-}(x)
\\
\varphi_{+}(x)
\end{pmatrix}
. \label{103}
\end{equation}
The new amplitudes $\varphi_{-}(x)$ and $\varphi_{+}(x)$ satisfy
decoupled differential equations, whose solutions are
\begin{equation}
\varphi_{\mp}(x) = \exp\!\left[ \pm i \int_{0}^{x} \text{d}x' \, \mu \, B_{\perp}(x') \right] \varphi_{\mp}(0) .
\label{104}
\end{equation}
If we consider an initial left-handed neutrino, we have
\begin{equation}
\begin{pmatrix}
\varphi_{L}(0)
\\
\varphi_{R}(0)
\end{pmatrix}
=
\begin{pmatrix}
1
\\
0
\end{pmatrix}
\quad \Rightarrow \quad
\begin{pmatrix}
\varphi_{-}(0)
\\
\varphi_{+}(0)
\end{pmatrix}
= \frac{1}{\sqrt{2}}
\begin{pmatrix}
1
\\
1
\end{pmatrix}
. \label{105}
\end{equation}
Then, the probability of $\nu_{L}\to\nu_{R}$ transitions is given
by
\begin{equation}
P_{\nu_{L}\to\nu_{R}}(x) = |\varphi_{R}(x)|^{2} = \sin^{2}\!\left( \int_{0}^{x} \text{d}x' \, \mu \, B_{\perp}(x') \right)
. \label{106}
\end{equation}
Note that the transition probability is independent from the
neutrino energy (contrary to the case of flavor oscillations) and
the amplitude of the oscillating probability is unity. Hence, when
the argument of the sine is equal to $\pi/2$ there is complete
$\nu_{L}\to\nu_{R}$ conversion.

The precession $\nu_{eL}\to\nu_{eR}$ in the magnetic field of the
Sun was considered in 1971 \cite{Cisneros:1971nq} as a possible
solution of the solar neutrino problem. If neutrinos are Dirac
particles, right-handed neutrinos are sterile and a
$\nu_{eL}\to\nu_{eR}$ conversion could explain the disappearance of
active solar $\nu_{eL}$'s.

In 1986 it was realized \cite{Voloshin:1986ty,Okun:1986na} that
the matter effect during neutrino propagation inside of the Sun
suppresses $\nu_{eL}\to\nu_{eR}$ transition by lifting the
degeneracy of $\nu_{eL}$ and $\nu_{eR}$. Indeed, taking into
account matter effects, the evolution equation~(\ref{102}) for a Dirac neutrino becomes
\begin{equation}
i \frac{ \text{d} }{ \text{d}x }
\begin{pmatrix}
\varphi_{L}(x)
\\
\varphi_{R}(x)
\end{pmatrix}
=
\begin{pmatrix}
V(x) & \mu B_{\perp}(x)
\\
\mu B_{\perp}(x) & 0
\end{pmatrix}
\begin{pmatrix}
\varphi_{L}(x)
\\
\varphi_{R}(x)
\end{pmatrix}
, \label{107}
\end{equation}
with the appropriate potential $V(x)$ which depends on the neutrino
flavor:
\begin{equation}
V_{\alpha}(x) = V_{\text{CC}}(x) \, \delta_{\alpha e} + V_{\text{NC}}(x)
. \label{i025}
\end{equation}
Here,
$V_{\text{CC}}$
and
$V_{\text{NC}}$
are the charged-current and neutral-current potentials given by
\begin{equation}
V_{\text{CC}}(x) = \sqrt{2} \, G_{\text{F}} \, N_{e}(x) , \qquad V_{\text{NC}}(x) = -
\frac{1}{2} \, \sqrt{2} \, G_{\text{F}} \, N_{n}(x) , \label{i017}
\end{equation}
where
$N_{e}(x)$ and $N_{n}(x)$ are
the electron and neutron number densities in the medium.
For antineutrinos,
$\overline{V}_{\alpha}(x) = - V_{\alpha}(x)$.

In the case of a
constant matter density, the differential equation (\ref{107}) can be solved
analytically with the orthogonal transformation
\begin{equation}
\begin{pmatrix}
\varphi_{L}(x)
\\
\varphi_{R}(x)
\end{pmatrix}
=
\begin{pmatrix}
\cos\xi & \sin\xi
\\
- \sin\xi & \cos\xi
\end{pmatrix}
\begin{pmatrix}
\varphi_{-}(x)
\\
\varphi_{+}(x)
\end{pmatrix}
. \label{108}
\end{equation}
The angle $\xi$ is chosen in order to diagonalize the matrix
operator in Eq.~(\ref{107}):
\begin{equation}
\sin 2 \xi = \frac{ 2 \mu B_{\perp} }{ \Delta{E}_{\text{M}} } ,
\label{109}
\end{equation}
with the effective energy splitting in matter
\begin{equation}
\Delta{E}_{\text{M}} = \sqrt{ V^{2} + \left( 2 \mu B_{\perp} \right)^{2} } . \label{109a}
\end{equation}
The decoupled evolution of $\varphi_{\mp}(x)$ is given by
\begin{equation}
\varphi_{\mp}(x) = \exp\!\left[ - \frac{i}{2} \left( V \mp \Delta{E}_{\text{M}} \right) \right] \varphi_{\mp}(0) .
\label{110}
\end{equation}
For an initial left-handed neutrino,
\begin{equation}
\begin{pmatrix}
\varphi_{-}(0)
\\
\varphi_{+}(0)
\end{pmatrix}
=
\begin{pmatrix}
\cos\xi
\\
\sin\xi
\end{pmatrix}
, \label{111}
\end{equation}
leading to the oscillatory transition probability
\begin{equation}
P_{\nu_{L}\to\nu_{R}}(x) = |\varphi_{R}(x)|^{2} = \sin^{2} 2 \xi
\sin^{2}\!\left( \frac{1}{2} \, \Delta{E}_{\text{M}} x \right) .
\label{1061}
\end{equation}
Since in matter $ \Delta{E}_{\text{M}}> 2 \mu B_{\perp}$, the
matter effect suppresses the amplitude of $\nu_{L}\to\nu_{R}$
transitions. However, these transitions are still independent from
the neutrino energy, which does not enter in the evolution
equation (\ref{107}).

When it was known, in 1986 \cite{Voloshin:1986ty,Okun:1986na},
that the matter potential has the effect of suppressing
$\nu_{L}\to\nu_{R}$ transitions because it breaks the degeneracy
of left-handed and right-handed states, it did not take long to
realize, in 1988
\cite{Akhmedov:1988nc,Lim:1988tk},
that the
matter potentials can cause resonant spin-flavor precession if
different flavor neutrinos have transition magnetic moments
(spin-flavor precession in vacuum was previously discussed in
\cite{Schechter:1981hw}).

Let us denote with
$\psi_{\alpha h}(x)$
the flavor and helicity amplitudes
(with $\alpha=e,\mu,\tau$ and $h=\pm1$),
i.e.
$\varphi_{\alpha -1}(x) \equiv \varphi_{\alpha L}(x)$
and
$\varphi_{\alpha +1}(x) \equiv \varphi_{\alpha R}(x)$.
Considering neutrino mixing,
the evolution of the flavor and helicity amplitudes
is given by
\begin{equation}
i \, \frac{\text{d}\psi_{\alpha h}(x)}{\text{d}x}
=
\sum_{\beta}
\sum_{h'=\pm1}
\left[
\left(
\sum_{k}
U_{\alpha k}
\frac{m_{k}^{2}}{2E}
U_{\beta k}^{*}
+
V_{\alpha}(x)
\delta_{\alpha\beta}
\right)
\delta_{hh'}
+
\mgm_{\alpha\beta}
B_{\perp}(x)
\delta_{-hh'}
\right]
\psi_{\beta h'}(x)
,
\label{p26}
\end{equation}
with the effective magnetic moments in the flavor basis
\begin{equation}
\mgm_{\alpha\beta}
=
\sum_{k,j}
U_{\alpha k}
\mgm_{kj}
U_{\beta j}^{*}
.
\label{p27}
\end{equation}

For a Dirac neutrino,
from Eq.~(\ref{A035a})
we have
\begin{equation}
\mgm_{jk} = {\mgm_{kj}}^{*}
\quad
\Rightarrow
\quad
\mgm_{\beta\alpha}
=
\mgm_{\alpha\beta}^{*}
.
\label{p28}
\end{equation}
If CP is conserved,
from Eq.~(\ref{A036}) and the reality of the mixing matrix,
for a Dirac neutrino
we obtain
\begin{equation}
\text{CP}
\quad
\Rightarrow
\quad
\mgm_{jk} = \mgm_{kj}
\quad
\Rightarrow
\quad
\mgm_{\beta\alpha}
=
\mgm_{\alpha\beta}
.
\label{p29}
\end{equation}

For a Majorana neutrino,
from Eqs.~(\ref{A044}) and (\ref{A046})
we have
\begin{equation}
\mgm_{jk} = - \mgm_{kj}
,
\qquad
\mgm_{kj} = - {\mgm_{kj}}^{*}
.
\label{p30}
\end{equation}
Hence,
in the mass basis of Majorana neutrinos there are no diagonal magnetic moments
and the transition magnetic moments are imaginary.
If CP is not conserved,
the mixing matrix is not real
and
the constraints (\ref{p30})
do not imply similar relations between the effective magnetic moments in the flavor basis,
for which we have only the relation in Eq.~(\ref{p28}) as for Dirac neutrinos.
In particular,
Majorana neutrinos can have diagonal effective magnetic moments in the flavor basis
if CP is not conserved.
Let us emphasize that both Dirac and Majorana phases contribute to this effect.
Therefore, it occurs also in the case of two-neutrino mixing,
in which there is one Majorana phase.

On the other hand,
if CP is conserved,
there is no additional constraint on the magnetic moments of Majorana neutrinos in the mass basis,
as we have seen in Section~\ref{ff:Majorana}.
However, in this case the mixing matrix is real and we have
\begin{equation}
\text{CP}
\quad
\Rightarrow
\quad
\mgm_{\beta\alpha}
=
-
\mgm_{\alpha\beta}
,
\qquad
\mgm_{\alpha\beta} = - {\mgm_{\alpha\beta}}^{*}
.
\label{p31}
\end{equation}
Hence, only if CP is conserved there are no diagonal magnetic moments
of Majorana neutrinos in the flavor basis
as in the mass basis.

In the following we discuss the spin-flavor evolution equation
in the two-neutrino mixing approximation,
which is interesting for understanding the relevant features of
neutrino spin-flavor precession.

Considering Dirac neutrinos,
from Eq.~(\ref{p26}) it follows that
the generalization of Eq.~(\ref{102})
to two-neutrino $\nu_{e}$--$\nu_{\mu}$ mixing
is, using an analogous notation,
\begin{equation}
i \frac{ \text{d} }{ \text{d}x }
\begin{pmatrix}
\varphi_{eL}(x)
\\
\varphi_{\mu L}(x)
\\
\varphi_{eR}(x)
\\
\varphi_{\mu R}(x)
\end{pmatrix}
=
\mathrm{H}
\begin{pmatrix}
\varphi_{eL}(x)
\\
\varphi_{\mu L}(x)
\\
\varphi_{eR}(x)
\\
\varphi_{\mu R}(x)
\end{pmatrix}
, \label{202}
\end{equation}
with the effective Hamiltonian matrix
\begin{equation}
\mathrm{H} =
\begin{pmatrix}
- \frac{ \Delta{m}^{2} }{ 4 E } \cos{2\vartheta} + V_{e} & \frac{ \Delta{m}^{2} }{ 4 E } \sin{2\vartheta} & \mu_{ee}
B_{\perp}(x) & \mu_{e\mu} B_{\perp}(x)
\\
\frac{ \Delta{m}^{2} }{ 4 E } \sin{2\vartheta} & \frac{ \Delta{m}^{2} }{ 4 E } \cos{2\vartheta} + V_{\mu} & \mu_{e\mu}^{*}
B_{\perp}(x) & \mu_{\mu\mu} B_{\perp}(x)
\\
\mu_{ee} B_{\perp}(x) & \mu_{e\mu} B_{\perp}(x) & - \frac{ \Delta{m}^{2} }{ 4 E } \cos{2\vartheta} & \frac{
\Delta{m}^{2} }{ 4 E } \sin{2\vartheta}
\\
\mu_{e\mu}^{*} B_{\perp}(x) & \mu_{\mu\mu} B_{\perp}(x) & \frac{ \Delta{m}^{2} }{ 4 E } \sin{2\vartheta} & \frac{
\Delta{m}^{2} }{ 4 E } \cos{2\vartheta}
\end{pmatrix}
, \label{203}
\end{equation}
where we have used the constraint (\ref{p28}) for the transition magnetic moments.
The matter potential can generate resonances,
which occur when two diagonal
elements of $\mathrm{H}$ become equal.
Besides the standard MSW resonance in the $\nu_{eL}\leftrightarrows\nu_{\mu L}$ channel
for
$
V_{\text{CC}} = \Delta{m}^{2} \cos2\vartheta / 2 E
$
(see
\cite{Giunti-Kim-2007,GonzalezGarcia:2007ib,Bilenky:2010zza,Xing:2011zza}),
there are two possibilities:
\begin{enumerate}
\item
There is a resonance in the
$\nu_{eL}\leftrightarrows\nu_{\mu R}$
channel for
\begin{equation}
V_{e} = \frac{ \Delta{m}^{2} }{ 2 E } \, \cos2\vartheta
.
\label{211}
\end{equation}
The density at which this resonance occurs is not the same as that of the MSW resonance,
because of the neutral-current contribution to $V_{e}=V_{\text{CC}}+V_{\text{NC}}$.
The location of this resonance depends on both $N_{e}$ and $N_{n}$.
\item There is a resonance in the $\nu_{\mu L}\leftrightarrows\nu_{eR}$ channel for
\begin{equation}
V_{\mu} = - \frac{ \Delta{m}^{2} }{ 2 E } \, \cos2\vartheta
.
\label{212}
\end{equation}
If $\cos2\vartheta>0$, this resonance is possible in normal matter, since the sign of $V_{\mu}=V_{\text{NC}}$ is
negative, as one can see from Eq.~(\ref{i017}).
\end{enumerate}
In practice the effect of these resonances could be the disappearance
of active $\nu_{eL}$ or $\nu_{\mu L}$ into sterile right-handed
states.

Let us consider now the more interesting case of Majorana neutrinos, which presents two fundamental differences with
respect to the Dirac case:
\begin{enumerate}
\renewcommand{\labelenumi}{\theenumi}
\renewcommand{\theenumi}{(\Alph{enumi})}
\item If CP is conserved Majorana neutrinos can have only a transition magnetic moment $\mu_{e\mu}=-\mu_{\mu e}=-\mu_{e\mu}^{*}$ in the flavor basis.
\item The right-handed states are not sterile, but interact as right-handed Dirac antineutrinos.
\end{enumerate}
Assuming CP conservation,
the evolution equation of the amplitudes is given by Eq.~(\ref{202}) with the effective Hamiltonian matrix
\begin{equation}
\mathrm{H} =
\begin{pmatrix}
- \frac{ \Delta{m}^{2} }{ 4 E } \cos{2\vartheta} + V_{e} & \frac{ \Delta{m}^{2} }{ 4 E } \sin{2\vartheta} & 0 &
\mu_{e\mu} B_{\perp}(x)
\\
\frac{ \Delta{m}^{2} }{ 4 E } \sin{2\vartheta} & \frac{ \Delta{m}^{2} }{ 4 E } \cos{2\vartheta} + V_{\mu} & -
\mu_{e\mu} B_{\perp}(x) & 0
\\
0 & \mu_{e\mu} B_{\perp}(x) & - \frac{ \Delta{m}^{2} }{ 4 E } \cos{2\vartheta} - V_{e} & \frac{ \Delta{m}^{2} }{ 4 E
} \sin{2\vartheta}
\\
-\mu_{e\mu} B_{\perp}(x) & 0 & \frac{ \Delta{m}^{2} }{ 4 E } \sin{2\vartheta} & \frac{ \Delta{m}^{2} }{ 4 E }
\cos{2\vartheta} - V_{\mu}
\end{pmatrix}
. \label{303}
\end{equation}
Again,
besides the standard MSW resonance in the $\nu_{eL}\leftrightarrows\nu_{\mu L}$ channel,
there are two possible resonances:
\begin{enumerate}
\item There is a resonance in the $\nu_{eL}\leftrightarrows\nu_{\mu R}$ channel for
\begin{equation}
V_{\text{CC}}+2V_{\text{NC}} = \frac{ \Delta{m}^{2} }{ 2 E } \, \cos2\vartheta
.
\label{311}
\end{equation}
\item There is a resonance in the $\nu_{\mu L}\leftrightarrows\nu_{eR}$ channel for
\begin{equation}
V_{\text{CC}}+2V_{\text{NC}} = - \frac{ \Delta{m}^{2} }{ 2 E } \, \cos2\vartheta
.
\label{312}
\end{equation}
\end{enumerate}
The location of both resonances depend on both $N_{e}$ and $N_{n}$. If $\cos2\vartheta>0$, only the first resonance can
occur in normal matter, where $ N_{n} \simeq N_{e}/6 $. A realization of the second resonance requires a large neutron
number density, as that in a neutron star.

The neutrino spin oscillations in a transverse magnetic field with
a possible rotation of the field-strength vector in a plane
orthogonal to the neutrino-propagation direction (such rotating
fields may exist in the convective zone of the Sun) have been
considered in
\cite{Vidal:1990fr,Smirnov:1991ia,Akhmedov:1993sh,Likhachev:1990ki}.
The effect of the magnetic-field rotation may substantially shift
the resonance point of neutrino oscillations. Neutrino spin
oscillations in electromagnetic fields with other different
configurations, including a longitudinal magnetic field and the
field of an electromagnetic wave, were examined in
\cite{Akhmedov:1988hd,Akhmedov:1990ng} and
\cite{Egorov:1999ah,Lobanov:2001ar,Dvornikov:2001ez,Dvornikov:2004en}.

It is possible to formulate a criterion \cite{Likhachev:1990ki}
for finding out if the neutrino spin and spin-flavor precession is
significant for given neutrino and background medium properties.
The probability of oscillatory transitions between two neutrino
states $\nu_{\alpha L}\leftrightarrows\nu_{\beta R}$ can be
expressed in terms of the elements of the effective Hamiltonian
matrices (\ref{203}) and (\ref{303}) as
\begin{equation}
P_{\nu_{\alpha L} \leftrightarrows \nu_{\beta R}}=\sin^{2} \vartheta_{\text{eff}} \sin^{2} \frac{x\pi}{L_{\text{eff}}},
\end{equation}
where
\begin{equation}
\sin^{2} \vartheta_{\text{eff}}=
\frac{4\mathrm{H}^{2}_{\alpha\beta}}{4\mathrm{H}^{2}_{\alpha\beta}+(\mathrm{H}_{\beta\beta}-\mathrm{H}_{\alpha\alpha})^{2}}
,
\qquad
L_{\text{eff}}=\frac {2\pi}{\sqrt
{4\mathrm{H}_{\alpha\beta}^{2}+(\mathrm{H}_{\beta\beta}-\mathrm{H}_{\alpha\alpha})^{2}}}
.
\label{leff}
\end{equation}
The transition probability can be of order unity
if the following two conditions hold simultaneously:
1) the amplitude
of the transition probability must be sizable (at least
$\sin^{2} \vartheta_{\text{eff}} \gtrsim 1/2$);
2) the neutrino path length in a
medium with a magnetic field should be longer than half the effective
length of oscillations $L_{\text{eff}}$. In accordance with this criterion,
it is possible to introduce the critical strength of a magnetic field
$B_{\text{cr}}$ which determines the region of field values
$B_{\perp}>B_{\text{cr}}$ at which the probability amplitude is not small ($\sin^{2}
\vartheta_{\text{eff}} > 1/2$):
\begin{equation}\label{B_cr}
B_{\text{cr}}=\frac {1}{2 {\tilde \mu}}\sqrt
{(\mathrm{H}_{\beta\beta}-\mathrm{H}_{\alpha\alpha})^{2}},
\end{equation}
where $\tilde \mu $ is $\mu_{ee}$, $\mu_{\mu \mu}$, $\mu_{e\mu}$, or
$\mu_{\mu e}$ depending on the type of neutrino transition process in question.

Consider, for instance, the case of
$\nu_{eL}\leftrightarrows\nu_{\mu R}$
transitions of Majorana neutrinos.
From Eqs.~(\ref{303}) and (\ref{B_cr}), it
follows \cite{Likhachev:1990ki} that
\begin{equation}
B_{\text{cr}}
=
\left|
\frac{1}{2\tilde\mu}
\left(
\frac{\Delta{m}^{2}\cos2\vartheta}{2E}
-
\sqrt{2}G_{\text{F}} N_{\text{eff}}
\right)
\right|,
\label{B_cr2}
\end{equation}
where $N_{\text{eff}}=N_{e}-N_n$.
For getting numerical estimates of $B_{\text{cr}}$ it is convenient
to re-write Eq.~(\ref{B_cr2}) in the following form:
\begin{equation}
B_{\text{cr}}
\approx
43 \, \frac{\mu_{B}}{\tilde \mu}
\left|
\cos2\vartheta
\left(\frac{\Delta{m}^{2}}{\text{eV}^{2}}\right)
\left(\frac{\text{MeV}}{E}\right)
-
2.5\times 10^{-31}
\left(\frac{N_{\text{eff}}}{\text{cm}^{-3}}\right)
\right|
\text{Gauss}
.
\label{cr3}
\end{equation}

An interesting feature of the evolution equation~(\ref{202}) in the case of
Majorana neutrinos is that the interplay of spin precession and flavor
oscillations can generate $\nu_{eL}\to\nu_{eR}$ transitions
\cite{Akhmedov:1991uk}. Since $\nu_{eR}$ interacts as right-handed Dirac
antineutrinos, it is often denoted by $\bar\nu_{eR}$, or only $\bar\nu_{e}$,
and called ``electron antineutrino''. This state can be detected through the
inverse $\beta$-decay reaction
\begin{equation}
\bar\nu_{e} + p \to n + e^{+} , \label{g096}
\end{equation}
having a threshold $ E_{\text{th}} = 1.8 \, \text{MeV} $.

The possibility of $\nu_{e}\to\bar\nu_{e}$ transitions
generated by spin-flavor precession of Majorana neutrinos is particularly interesting
for solar neutrinos, which experience matter effects in the
interior of the Sun in the presence of the solar magnetic field
(see \cite{Pulido:1991fb,Shi:1992ek}).
Taking into account
the dominant $\nu_{e}\to\nu_{a}$
transitions due to neutrino oscillations
(see \cite{Giunti-Kim-2007,GonzalezGarcia:2007ib,Bilenky:2010zza,Xing:2011zza}),
with
$\nu_{a}=\cos\vartheta_{23}\nu_{\mu}-\sin\vartheta_{23}\nu_{\tau}$
and
$\sin^2 2\vartheta_{23} > 0.95$ (90\% C.L.)
\cite{PDG-2012},
the probability of solar $\nu_{e}\to\bar\nu_{e}$ transitions
is given by
\cite{Akhmedov:2002mf}
\begin{equation}
P_{\nu_{e}\to\bar\nu_{e}}
\simeq
1.8 \times 10^{-10}
\sin^2 2\vartheta_{12}
\left(
\frac{\mu_{ea}}{10^{-12}\,\mu_{B}}
\,
\frac{B_{\perp}(0.05R_{\odot})}{10\,\text{kG}}
\right)^2
,
\label{psun}
\end{equation}
where $\mu_{ea}$ is the transition magnetic moment between
$\nu_{e}$ and $\nu_{a}$,
$\sin^2 2\vartheta_{12} = 0.857^{+0.023}_{-0.025}$
\cite{PDG-2012}
and
$R_{\odot}$
is the radius of the Sun.

It is also possible that spin-flavor precession
occurs in the convective zone of the Sun,
where there can be random turbulent magnetic fields
\cite{Miranda:2003yh,Miranda:2004nz,Friedland:2005xh}.
In this case
\cite{Raffelt:2009mm},
\begin{align}
P_{\nu_{e}\to\bar\nu_{e}}
\approx
\null & \null
10^{-7} S^2
\left( \frac{\mu_{ea}}{10^{-12}\,\mu_{B}} \right)^2
\left( \frac{B}{20\,\text{kG}} \right)^2
\left( \frac{3\times10^{4}\,\text{km}}{L_{\text{max}}} \right)^{p-1}
\nonumber
\\
\null & \null
\times
\left( \frac{8\times10^{-5}\,\text{eV}^2}{\Delta{m}^{2}_{21}} \right)^{p}
\left( \frac{E}{10\,\text{MeV}} \right)^{p}
\left( \frac{\cos^2\vartheta_{12}}{0.7} \right)^{p}
,
\label{psunturb}
\end{align}
where
$S$ is a factor of order unity depending on the spatial configuration of the magnetic field,
$B$ is the average strength of the magnetic field at the spatial scale $L_{\text{max}}$,
which is the largest scale of the turbulence,
$p$ is the power of the turbulence scaling,
$\Delta{m}^{2}_{21} = 7.50^{+0.19}_{-0.20}\times10^{-5}\,\text{eV}^2$ \cite{PDG-2012},
and
$E$ is the neutrino energy.
A possible value of $p$ is 5/3
\cite{Miranda:2003yh,Miranda:2004nz,Friedland:2005xh},
corresponding to Kolmogorov turbulence.
Conservative values for the other parameters are
$B=20\,\text{kG}$
and
$L_{\text{max}}=3\times10^{4}\,\text{km}$.

In 2002, the Super-Kamiokande Collaboration established for the
flux of solar $\bar\nu_{e}$'s a 90\% C.L. an upper limit of 0.8\%
of the Standard Solar Model (SSM) neutrino flux in the range of
energy from 8 to 20 MeV \cite{Gando:2002ub}
by taking as a reference the BP00 SSM prediction
$\phi_{^{8}\text{B}}^{\text{BP00}} = 5.05\times10^{6}\,\text{cm}^{-2}\,\text{s}^{-1}$
for the solar $^{8}\text{B}$ flux
\cite{Bahcall:2000nu}
and
assuming an undistorted $^{8}\text{B}$ spectrum for the $\bar\nu_{e}$'s.
This limit was
improved in 2003 by the KamLAND Collaboration
\cite{Eguchi:2003gg}
to
$ 2.8 \times 10^{-4} $
of the BP00 SSM prediction
at 90\% C.L.
by measuring
$\phi_{\bar\nu_{e}} < 370\,\text{cm}^{-2}\,\text{s}^{-1}$ (90\% C.L.)
in the energy range 8.3 -- 14.8 MeV,
which corresponds to
$\phi_{\bar\nu_{e}} < 1250\,\text{cm}^{-2}\,\text{s}^{-1}$ (90\% C.L.)
in the entire $^{8}\text{B}$ energy range
assuming an undistorted spectrum.

Recently,
the Borexino collaboration established the best limit on the probability of solar
$\nu_{e}\to\bar\nu_{e}$ transitions
\cite{Bellini:2010gn},
\begin{equation}
P_{\nu_{e}\to\bar\nu_{e}}
<
1.3 \times 10^{-4}
\qquad
\text{(90\% C.L.)}
,
\label{Borexino-antinue}
\end{equation}
by taking
as a reference
$\phi_{^{8}\text{B}}^{\text{SSM}} = 5.88\times10^{6}\,\text{cm}^{-2}\,\text{s}^{-1}$
\cite{Serenelli:2009yc}
and
assuming an undistorted $^{8}\text{B}$ spectrum for the $\bar\nu_{e}$'s.
They measured
$ \phi_{\bar\nu_{e}} < 320 \, \text{cm}^{-2} \, \text{s}^{-1} $ (90\% C.L.)
for $ E_{\bar\nu_{e}} > 7.3 \, \text{MeV} $,
which corresponds to
$ \phi_{\bar\nu_{e}} < 760 \, \text{cm}^{-2} \, \text{s}^{-1} $ (90\% C.L.)
in the entire $^{8}\text{B}$ energy range
assuming an undistorted spectrum

The implications of the limits
on the flux of solar $\bar\nu_{e}$'s on Earth
for the
spin-flavor precession of solar neutrinos have been studied in
several papers
\cite{Akhmedov:2002mf,Chauhan:2003wr,Miranda:2003yh,Miranda:2004nz,Balantekin:2004tk,Guzzo:2005rr,Friedland:2005xh,Yilmaz:2008vh},
taking into account the dominant $\nu_{e}\to\nu_{\mu},\nu_{\tau}$
transitions due to neutrino oscillations
(see \cite{Giunti-Kim-2007,GonzalezGarcia:2007ib,Bilenky:2010zza,Xing:2011zza}).
Using Eqs.~(\ref{psun}) and (\ref{Borexino-antinue}),
we obtain
\begin{equation}
\mu_{ea}
\lesssim
1.3 \times 10^{-12}
\,
\frac{7\,\text{MG}}{B_{\perp}(0.05R_{\odot})}
\,
\mu_{\text{B}}
,
\label{331a}
\end{equation}
with
$ 600 \, \text{G} \lesssim B_{\perp}(0.05R_{\odot}) \lesssim 7 \, \text{MG} $
\cite{Bellini:2010gn}.
In the case of spin-flavor precession in the convective zone of the Sun
with random turbulent magnetic fields,
Eqs.~(\ref{psunturb}) and (\ref{Borexino-antinue})
give,
assuming $p=5/3$,
\begin{equation}
\mu_{ea}
\lesssim
4 \times 10^{-11}
\,
S^{-1}
\,
\frac{20\,\text{kG}}{B}
\,
\left( \frac{L_{\text{max}}}{3\times10^{4}\,\text{km}} \right)^{1/3}
\,
\mu_{\text{B}}
.
\label{331b}
\end{equation}

The spin-flavor mechanism was also considered \cite{Pulido:2005pt} in
order to describe time variations of solar-neutrino fluxes in
Gallium experiments. The effect of a nonzero neutrino magnetic
moment is also of interest in connection with the analysis of
helioseismological observations \cite{Couvidat:2003ba}.

The idea that the neutrino magnetic moment may solve the supernova
problem, i.e. that the neutrino spin-flip transitions in a
magnetic field provide an efficient mechanism of energy transfer
from a protoneutron star, was first discussed in \cite{Dar:1987yv}
and then investigated in some detail in
\cite{Nussinov:1987zr,Goldman:1987fg,Lattimer:1988mf}.
The possibility of a
loss of up to half of the active left-handed neutrinos because of
their transition to sterile right-handed neutrinos in strong
magnetic fields at the boundary of the neutron star (the so-called
boundary effect) was considered in \cite{Likhachev:1990ki}.
\section{Summary and perspectives}
\label{sec8}

In this review we have considered the electromagnetic properties of
neutrinos with focus on the most important issues related to the
problem. The main results discussed in the paper can be summed up as
follows.

In the most general case, the neutrino electromagnetic vertex
function is defined in terms of four form factors:
the charge, dipole
magnetic and electric and anapole form factors. This decomposition
is consistent with Lorentz and electromagnetic gauge invariance.
The four form factors at zero momentum transfer $q^2=0$
are, respectively, the neutrino charge, magnetic moment, electric
moment and anapole moment. These quantities contribute to
elements of the scattering matrix and describe neutrino
interactions with real photons.

An important characteristic of neutrino electromagnetic
properties is that they are different for Dirac and Majorana neutrinos.
In particular, Majorana neutrinos cannot
have diagonal magnetic or electric moments.
Thus, studies of neutrino
electromagnetic interactions can be used as a procedure to
distinguish whether a neutrino is a Dirac or Majorana particle.

Moreover,
CP invariance in the lepton sector puts additional constraints on the neutrino form factors
and can be tested with experimental probes of
neutrino electromagnetic interactions.

Up to now,
no effect of neutrino electromagnetic properties
has been found in
terrestrial laboratory experiments and
in the analyses of
astrophysical and cosmological data.
However, massive neutrinos have non-trivial
electromagnetic properties in a wide set of theoretical
frameworks, including the simplest extension of the Standard Model
with inclusion of singlet right-handed neutrinos.
Therefore,
the search for non-vanishing neutrino electromagnetic properties
is of great interest for
experimentalist and theorists.

The neutrino dipole magnetic (and also electric) moment is
theoretically the most well studied and understood among the
neutrino electromagnetic moments.
In models which extend the Standard Model with the addition of singlet right-handed neutrinos,
a Dirac neutrino has a non-zero magnetic moment proportional to the
neutrino mass, that yields a very small value for the magnetic moment,
less than about
$3 \times 10^{-19} \, \mu_{B}$
for a neutrino mass smaller than 1 eV
(Eq.~(\ref{mu_3_10_19})). Extra terms contributing to
the magnetic moment of the neutrino that are not proportional to
the neutrino mass may exist, for example, in the framework of left-right symmetric
models. In this type of models, as well as in other
generalizations of the Standard Model, as for instance in
supersymmetric and extra-dimension models, values of
the neutrino magnetic moment
much larger than $10^{-19} \, \mu_{B}$ can be obtained.

The most severe terrestrial experimental upper bound for the
effective electron antineutrino magnetic moment,
$\mu_{{\bar {\nu}}_e} \leq 2.9 \times 10^{-11} \mu_{B}$ at 90\% C.L.
(Eq.~(\ref{lim-GEMMA})),
has been recently obtained in the direct
antineutrino-electron scattering experiment
performed by the GEMMA collaboration
\cite{Beda:2012-AHEP}.
This value is still about an order of
magnitude weaker than the upper bound
$\mu_{\nu} \lesssim 3 \times 10^{-12} \, \mu_{B}$
(Eq.~(\ref{p02}))
obtained from astrophysics (considering the cooling of red giant stars)
\cite{Raffelt:1990pj}.

There is a gap of some orders of magnitude between the
present experimental limits $\sim 10^{-11} \div 10^{-12} \, \mu_{B}$ on
neutrino magnetic moments and the predictions of different
extensions of the Standard Model which hint at a range
$\sim 10^{-14} \div 10^{-15} \, \mu_{B}$
\cite{Bell:2005kz,Bell:2006wi,Bell:2007nu}
(see Section~(\ref{sec:Theoretical})).
The terrestrial
experimental constraints have been improved by only one order of
magnitude during a period of about twenty years.
Further improvements are very important,
but unfortunately at the moment there is no new idea which could
lead to fast improvements in the near future.
On the other hand,
astrophysical studies could allow significant improvements
of the sensitivity to non-trivial neutrino
electromagnetic properties
and maybe find a positive indication in their favor.
In particular,
neutrino flows in extreme astrophysical environments with very strong
magnetic fields are sensitive to small values of the neutrino electromagnetic moments.
An example is the modelling of neutrinos
propagation during core-collapse supernovae
(see \cite{Pehlivan:2011hp} and references therein)
where very strong magnetic fields are believed to exist
and in which the
influence of neutrino electromagnetic properties has not yet been taken into account.

\section*{Acknowledgments}

The work of A. Studenikin on this paper has been partially supported by the Russian Foundation for Basic Research (grants N.~11-02-01509 and N.~12-02-06833) and the Ministry of Education and Science of Russia (state contract N.~12.741.11.0180 and projects N.~2012-1.2.1-12-000-1012-1958 and N.~2012-1.1-12-000-1011-6097).
C. Giunti would like to thank
the Department of Physics of the University of Torino
for hospitality and support.

%\bibliographystyle{h-physrev4}
%\bibliography{ahep}
%\bibliography{bibtex/nu}

\end{document}